\documentstyle[preprint,eqsecnum,aps]{revtex}
\draft
\begin{document}
\tighten
\title{Gravitational radiation reaction and balance equations\\
to post-Newtonian order}
\author{Luc Blanchet}
\address{D\'epartement d'Astrophysique Relativiste et de Cosmologie,\\
Centre National de la Recherche Scientifique (UPR 176),\\
Observatoire de Paris, 92195 Meudon Cedex, France}
\date{\today}  
	
\maketitle
\begin{abstract}
Gravitational radiation reaction forces and balance equations for energy
and momenta are investigated to 3/2 post-Newtonian (1.5PN) order beyond
the quadrupole approximation, corresponding to the 4PN order in the
equations of motion of an isolated system.  By matching a post-Newtonian
solution for the gravitational field inside the system to a
post-Minkowskian solution (obtained in a previous work \cite{B93}) for
the gravitational field exterior to the system, we determine the 1PN
relativistic corrections to the ``Newtonian" radiation reaction
potential of Burke and Thorne. The 1PN reaction potential involves both scalar
and vectorial components, with the scalar component depending on the
mass-type quadrupole and octupole moments of the system, and the
vectorial component depending in particular on the current-type
quadrupole moment.  In the case of binary systems, the 1PN radiation
reaction potential has been shown \cite{IW93,IW95} to yield consistent
results for the 3.5PN approximation in the binary's equations of motion.
Adding up the effects of tails, the radiation reaction is then written
to 1.5PN order.
In this paper, we establish the validity to 1.5PN order, for general
systems, of the balance equations relating the losses of energy, linear
momentum, and angular momentum in the system to the corresponding fluxes
in the radiation field far from the system.
\end{abstract}

\section{Introduction}  \label{sec:1}

The old idea (in any field theory) of losses of energy and momenta in an
isolated system, due to the presence of radiation reaction forces in the
equations of motion, is of topical interest in the case of the
gravitational field.  Notably, gravitational radiation reaction forces
play an important role in astrophysical binary systems of compact
objects (neutron stars or black holes).  The electromagnetic-based
observations of the Hulse-Taylor \cite{HT75} binary pulsar PSR~1913+16
have yielded evidence that the binding energy of the pulsar and its
companion decreases because of gravitational radiation reaction
\cite{TFMc79,D83b,TWoDW,T93}.

Even more relevant to the problem of radiation reaction are the future
gravitational-based observations of inspiralling (and then coalescing)
compact binaries.  The dynamics of these systems is entirely driven by
gravitational radiation reaction forces.  The future detectors such as
LIGO and VIRGO should observe the gravitational waves emitted during the
terminal phase, starting about twenty thousands orbital rotations before the
coalescence of two neutron stars.  Because inspiralling compact binaries
are very relativistic, and thanks to the large number of observed
rotations, the output of the detectors should be compared with a very
precise expectation from general relativity \cite{3mn,FCh93,CF94}.  In
particular Cutler {\it et al} \cite{3mn} have shown that our {\it a
priori} knowledge of the relativistic (or post-Newtonian) corrections in
the radiation reaction forces will play a crucial role in our ability to
satisfactorily extract information from the gravitational signals.
Basically, the reaction forces inflect the time evolution of the binary's
orbital phase, which can be determined very precisely because of the
accumulation of observed rotations.  The theoretical problem of the
phase evolution has been addressed using black-hole perturbation
techniques, valid when the mass of one body is small as compared with
the other mass \cite{P93,CFPS93,TNaka94,Sasa94,TSasa94,P95}, and using the
post-Newtonian theory, valid for arbitrary mass ratios
\cite{BDI95,BDIWW95,WWi96,B96pn}. It has been shown \cite{CFPS93,TNaka94,P95}
that post-Newtonian corrections in the radiation reaction forces should
be known up to at least the third post-Newtonian (3PN) order, or
relative order $c^{-6}$ in the velocity of light.

The radiation reaction forces in the equations of motion of a self-gravitating
system arise at the 2.5PN order (or $c^{-5}$ order) beyond the Newtonian
acceleration.  Controlling the $n$th post-Newtonian corrections in the
reaction force means, therefore, controlling the $(n+2.5)$th
post-Newtonian corrections in the equations of motion.  If $n=3$ this is
very demanding, and beyond our present knowledge.  A way out of this
problem is to {\it assume} the validity of a balance equation for
energy, which permits relating the mechanical energy loss in the system to the
corresponding flux of radiation far from the system.  Using such a
balance equation necessitates the knowledge of the equations of motion
up to the $n$PN order instead of the $(n+2.5)$PN one.  The price to be
paid for this saving is the computation of the far-zone flux up to the
same (relative) $n$PN order.  However, this is in general less demanding
than going to $(n+2.5)$PN order in the equations of motion.  All the
theoretical works on inspiralling binaries compute the phase evolution
from the energy balance equation.  Black-hole perturbations
\cite{TNaka94,Sasa94,TSasa94} reach $n=4$ in this way, and the
post-Newtonian theory \cite{BDI95,BDIWW95,WWi96,B96pn} has $n=2.5$.

An important theoretical problem is therefore to improve the present
situation by showing the validity to post-Newtonian order of the balance
equations for energy, and also for linear and angular momenta. This problem
is equivalent to controlling the radiation
reaction forces at the same post-Newtonian order.  Arguably,
this problem is also important in its own (not only for applications to
inspiralling compact binaries).

Radiation reaction forces in general relativity have long been investigated
(see \cite{D83a} for a review of works prior the seventies). In
the late sixties, Burke and Thorne \cite{Bu69,Th69,Bu71}, using a method
of matched asymptotic expansions, introduced a quasi-Newtonian
reactive potential, proportional to the fifth time-derivative of the
Newtonian quadrupole moment of the source.  At about the same time,
Chandrasekhar and collaborators \cite{C69,CN69,CE70}, pursuing a
systematic post-Newtonian expansion in the case of extended fluid
systems, found some reactive terms in the equations of motion at the
2.5PN approximation.  The reactive forces are different in the two
approaches because of the use of different coordinate systems, but both
yield secular losses of energy and angular momentum in
agreement with the standard Einstein quadrupole formulas (see however
\cite{EhlRGH,WalW80,BD84}). These results, after later confirmation and
improvements \cite{AD75,PaL81,Ehl80,Ker80,Ker80',BRu81,BRu82,S85},
show the validity of the balance equations to {\it Newtonian} order, and
in the case of weakly self-gravitating fluid systems.  In the case of
binary systems of compact objects (such as the binary pulsar and
inspiralling binaries), the Newtonian balance equations are also known to be
valid, as the complete dynamics of these systems
has been worked out (by Damour and Deruelle \cite{DD81a,DD81b,D82,D83a})
up to the 2.5PN order where radiation reaction effects appear.

Post-Newtonian corrections in the radiation reaction force can be obtained
from first principles using a combination of analytic approximation
methods.  The methods are (i) a post-Minkowskian or non-linear expansion
method for the field in the weak-field domain of the source (including
the regions far from the source), (ii) a multipolar expansion method for
each coefficient of the post-Minkowskian expansion in the domain
exterior to the source, and (iii) a post-Newtonian expansion method (or
expansion when $c\to\infty$) in the near-zone of the source (including its
interior).  Then an
asymptotic matching (in the spirit of \cite{Bu69,Th69,Bu71}) permits
us to connect the external field to the field inside the source.  Notably, the
methods (i) and (ii) have been developed by Blanchet and Damour
\cite{BD86,B87,BD92} on foundations laid by Bonnor and collaborators
\cite{Bo59,BoR66,HR69}, and Thorne \cite{Th80}.  The method (iii) and
matching have also been developed within the present approach
\cite{BD89,DI91a,B95}.

The post-Newtonian correction that is due to gravitational wave tails in
the reaction force was determined first using the latter methods
\cite{BD88}.  The tails of waves are produced by scattering of the
linear waves off the static spacetime curvature generated by the total
mass of the source (see e.g.  \cite{BoR66,HR69}).  Tails appear as
non-local integrals, depending on the full past history of the system, and
modifying its present dynamics by a post-Newtonian correction of 1.5PN order 
in the radiation reaction force, corresponding to 4PN order in the
equations of motion \cite{BD88}. It has been shown in
\cite{BD92} that the tail contribution in the reaction force is such
that the balance equation for energy is verified for this particular
effect.  This is a strong indication that the balance equations are
actually valid beyond the Newtonian order (1.5PN order in this case).
For completeness we shall include this result in the present paper.

The methods (i)-(ii) have been implemented in [1]
in order to investigate systematically the occurence and structure of the
contributions in the exterior field which are expected to yield
radiation reaction effects (after application of the method (iii) and the
relevant matching).  The present paper is the direct continuation of
the paper [1], that we shall refer here to as paper I.

Working first within the linearized theory, we investigated in paper I
the ``antisymmetric" component of the exterior field, a solution of the
d'Alembertian equation composed of a retarded (multipolar) wave minus
the corresponding advanced wave.  Antisymmetric waves in the exterior
field are expected to yield radiation reaction effects in the
dynamics of the source.  Indeed, these waves change sign when we reverse
the condition of retarded potentials into the advanced condition (in the
linearized theory), and have the property of being regular all over the
source (when the radial coordinate $r \to 0$).  Thus, by matching, the
antisymmetric waves in the exterior field are necessarily
present in the interior field as well, and can be interpreted
as radiation reaction potentials. In a particular coordinate
system suited to the (exterior) near zone of the source (and constructed
in paper~I), the antisymmetric waves define a radiation reaction
{\it tensor} potential in the linearized theory, generalizing the
radiation reaction scalar potential of Burke and Thorne
\cite{Bu69,Th69,Bu71}.

Working to non-linear orders in the post-Minkowskian approximation, we
introduced in paper~I a particular decomposition of the retarded
integral into the sum of an ``instantaneous" integral, and an
homogeneous solution composed of antisymmetric waves (in the same
sense as in the linearized theory).  The latter waves are associated
with radiation reaction effects of non-linear origin.  For instance,
they contain the non-linear tail contribution obtained previously
\cite{BD88}.  At the 1PN order, the non-linear effects lead simply to
a re-definition of the multipole moments which parametrize the
linearized radiation reaction potential.  However, the radiation
reaction potential at 1PN order has been derived only in the external
field.  Thus, it was emphasized in paper~I that in order to
meaningfully interpret the physical effects of radiation reaction, it
is necessary to complete this derivation by an explicit matching to
the field inside the source.

We perform the relevant matching (at 1PN order) in the present paper.
Namely, we obtain a solution for the field inside the source
(satisfying the non-vacuum field equations), which can be transformed
by means of a suitable coordinate transformation (in the exterior
near-zone of the source) into the exterior field determined in paper
I.  The matching yields in particular the multipole moments
parametrizing the reaction potential as explicit integrals over the
matter fields in the source.  As the exterior field satisfies
physically sensible boundary conditions at infinity (viz the
no-incoming radiation condition imposed at past-null infinity), the
1PN-accurate radiation reaction potentials are, indeed, appropriate
for the description of the dynamics of an isolated system.

To 1PN order beyond the Burke-Thorne term, the reaction
potential involves a scalar potential, depending on the mass-type
quadrupole and octupole moments of the source, and a vectorial potential,
depending in particular on the current-type quadrupole moment.  The
existence of such vectorial component was first noticed in the
physically restricting case where the dominant quadrupolar radiation is
suppressed \cite{BD84}.

A different approach to the problem of radiation reaction has been
proposed by Iyer and Will \cite{IW93,IW95} in the case of binary systems
of point particles.  The expression of the radiation reaction force is
deduced, in this approach, from the {\it assumption} that the balance
equations for energy and angular momentum are correct (the angular
momentum balance equation being necessary for non-circular orbits).
Iyer and Will determine in this way the 2.5PN and 3.5PN approximations
in the equations of motion of the binary, up to exactly the freedom left
by the un-specified coordinate system.  They also check that the
1PN-accurate radiation reaction potentials of the present paper (and
paper~I) correspond in their formalism, when specialized to binary
systems, to a unique and consistent choice of a coordinate system. This
represents a non-trivial check of the validity of the
1PN reaction potentials.

In the present paper we prove that the 1PN-accurate radiation reaction
force in the equations of motion of a general system extracts energy,
linear momentum and angular momentum from the system at the same rate as
given by the (known) formulas for the corresponding radiation fluxes at
infinity. The result is extended to include the tails at 1.5PN order.
Thus we prove the validity up to the 1.5PN order of the energy and
momenta balance equations (which were previously known to hold at
Newtonian order, and for the specific effects of tails at 1.5PN order).

Of particular interest is the loss of linear momentum, which can be
viewed as a ``recoil'' of the center of mass of the source in reaction
to the wave emission.  This effect is purely due to the 1PN corrections
in the radiation reaction potential, and notably to its vectorial
component (the Newtonian reaction potential predicts no recoil). Numerous
authors have obtained this effect by computing the {\it flux} of linear
momentum at infinity, and then by relying on the balance equation to get
the actual recoil \cite{BoR61,Pa71,Bek73,Press77}.  Peres \cite{Peres62}
made a direct computation of the linear momentum loss in the source, but
limited to the case of the linearized theory. Here we prove
the balance equation for linear momentum in the full non-linear theory.

The results of this paper apply to a weakly self-gravitating system.
The case of a source made of strongly self-gravitating (compact)
objects is {\it a priori} excluded.  However, the theoretical works on
the Newtonian radiation reaction in the binary system PSR~1913+16
\cite{DD81a,DD81b,D82,D83a} have shown that some ``effacement'' of the
internal structure of the compact bodies is at work in general
relativity.  Furthermore, the computation of the radiation reaction at
1PN order in the case of two point-masses \cite{IW93,IW95} has shown
agreement with a formal reduction, by means of $\delta$-functions, of
the 1PN radiation reaction potentials, initially derived in this paper
only in the case of weakly self-gravitating systems.  These works give
us hope that the results of this paper will remain unchanged in the
case of systems containing compact objects. If this is the case, the
present derivation of the 1.5PN balance equations constitutes a clear
support of the usual way of computing the orbital phase evolution of
inspiralling compact binaries
\cite{3mn,FCh93,CF94,P93,CFPS93,TNaka94,Sasa94,TSasa94,P95,BDI95,BDIWW95,WWi96,B96pn}.

The plan of this paper is the following.  The next section (II) is
devoted to several recalls from paper I which are necessary in order
that the present paper be essentially self-contained.  In Section III
we obtain, using the matching procedure, the gravitational field
inside the source, including the 1PN reactive contributions. Finally,
in Section IV, we show that the latter reactive contributions, when
substituted into the local equations of motion of the source, yield
the expected 1PN and then 1.5PN balance equations for energy and
momenta.

\section{Time-asymmetric structure of gravitational radiation}\label{sec:2}

\subsection{Antisymmetric waves in the linearized metric}

Let $D_e=\{({\bf x},t), |{\bf x}|>r_e\}$ be the domain exterior to the
source, defined by $r_e>a$, where $a$ is the radius of the
source. We assume that the gravitational field is weak everywhere,
inside and outside the source. In particular $a\gg GM/c^2$, where $M$ is
the total mass of the source.  Let us consider, in $D_e$, the
gravitational field at the first approximation in a non-linearity
expansion.  We write the components $h^{\mu\nu}$ of the
deviation of the metric density from the Minkowski metric
$\eta^{\mu\nu}$ in the form \cite{N}

\begin{equation}
 h^{\mu\nu} \equiv \sqrt{-g} g^{\mu\nu} - \eta^{\mu\nu} =
    Gh^{\mu\nu}_{(1)} + O (G^2)\ , \label{eq:2.1}
\end{equation}
where the coefficient of the Newton constant $G$ represents the linearized
field $h_{(1)}^{\mu\nu}$, satisfying the vacuum
linearized field equations in $D_e$,
\begin{equation}
 \Box h^{\mu\nu}_{(1)} = \partial^\mu \partial_\lambda h^{\lambda\nu}_{(1)}
 + \partial^\nu \partial_\lambda h^{\lambda\mu}_{(1)}
 - \eta^{\mu\nu} \partial_\lambda \partial_\sigma h^{\lambda\sigma}_{(1)}
 \ .\label{eq:2.2}
\end{equation}
We denote by $\Box\equiv\eta^{\mu\nu}\partial_\mu\partial_\nu$ the
flat space-time d'Alembertian operator.

The general solution 
of the equations (\ref{eq:2.2}) in $D_e$ can be parametrized
(modulo an arbitrary linearized coordinate transformation)
by means of two and only two sets of multipole moments, referred to as the
mass-type moments, denoted $M_L$, and the current-type moments, $S_L$
\cite{Th80}. The capital letter $L$ represents a multi-index composed of
$l$ indices, $L=i_1i_2\cdots i_l$ (see \cite{N} for our notation
and conventions). The multipolarity of the moments is $l\geq 0$
in the case of the mass moments, and $l\geq 1$ in the case of the current
moments.  The $M_L$'s and $S_L$'s are symmetric and trace-free
(STF) with respect to their $l$
indices.  The lowest-order moments $M$, $M_i$ and $S_i$ are constant, and
equal respectively to the total constant mass (including the energy of the
radiation to be emitted), to the position of the center of
mass times the mass, and to the total angular momentum of the source.
The higher-order moments, having $l\geq 2$, are arbitrary functions of time,
$M_L(t)$ and $S_L(t)$, which encode all the physical properties of the
source as seen in the exterior (linearized) field.  In terms
of these multipole moments, the ``canonical'' linearized solution of
Thorne \cite{Th80} reads
\begin{mathletters}
\label{eq:2.3}
\begin{eqnarray}
 h^{00}_{\rm can(1)} &&= -{4 \over c^2}
  \sum_{l \geq 0}{(-)^{l} \over l !}\partial_ L
\left[{1\over r} M_L\left(t- {r \over c}\right)\right]\ ,\label{eq:2.3a}\\
  h^{0i}_{\rm can(1)} &&= {4 \over c^3} \sum_{l \geq 1} {(-)^{l} \over
   l !} \partial_{L-1} \left[ {1\over r} M^{(1)}_{iL-1} \left(t- {r
   \over c} \right) \right] \nonumber \\ && + {4 \over c^3} \sum_{l
   \geq 1} {(-)^{l} l \over (l +1)!} \varepsilon_{iab}\partial_{aL-1}
\left[ {1\over r} S_{bL-1} \left(t- {r\over c}\right)\right]\ ,
             \label{eq:2.3b} \\
  h^{ij}_{\rm can(1)} &&= - {4 \over c^4}
   \sum_{l \geq 2}{(-)^{l} \over l!}\partial_{L-2}
   \left[ {1\over r} M^{(2)}_{ijL-2}
   \left(t- {r \over c} \right) \right] \nonumber \\
  && - {8 \over c^4} \sum_{l \geq 2} {(-)^{l} l \over (l +1)!}
  \partial_{aL-2} \left[{1\over r}\varepsilon_{ab(i} S^{(1)}_{\!j)bL-2}
   \left(t- {r \over c} \right) \right] \ .  \label{eq:2.3c}
\end{eqnarray}
\end{mathletters}
This solution satisfies (\ref{eq:2.2}) and the condition of harmonic
coordinates (i.e. $\partial_\nu h^{\mu\nu}_{\rm can(1)} =0$).  Here we
impose that the multipole moments $M_L(t)$ and $S_L(t)$ are constant
in the remote past, before some finite instant $-\cal T$ in the
past. With this assumption the linearized field (\ref{eq:2.3}) (and
all the subsequent non-linear iterations built on it) is stationary in
a neigthborhood of past-null infinity and spatial infinity (it
satisfies time-asymmetric boundary conditions in space-time).  This
ensures that there is no radiation incoming on the system, which would
be produced by some sources located at infinity.

In paper I (Ref. \cite{B93}) the contribution in (\ref{eq:2.3}) which
changes sign when we reverse the condition of retarded potentials to the
advanced condition was investigated. This contribution is
obtained by replacing each retarded wave in (\ref{eq:2.3}) by the
corresponding antisymmetric wave, half
difference between the retarded wave and the corresponding advanced one.
The antisymmetric wave changes sign when we reverse the time evolution
of the moments $M_L(t)$ and $S_L(t)$, say $M_L(t)\to M_L(-t)$, and
evaluate afterwards the wave at the reversed time $-t$.
Thus, (\ref{eq:2.3}) is decomposed as
\begin{equation}
 h^{\mu\nu}_{{\rm can}(1)} =
 \left( h^{\mu\nu}_{{\rm can}(1)} \right)_{\rm sym} + \left(
 h^{\mu\nu}_{{\rm can}(1)} \right)_{\rm antisym}\ . \label{eq:2.4}
\end{equation}
The symmetric part is given by
\FL
\begin{mathletters}
\label{2.5}
\begin{eqnarray}
 \left( h^{00}_{{\rm can}(1)} \right)_{\rm sym} =&& - {4 \over c^2}
    \sum_{l \geq 0} {(-)^l \over l!}
    \partial_L \left\{{ M_L (t-r/c) + M_L(t+r/c) \over 2r }\right\}
    \ , \label{eq:2.5a} \\
 \left( h^{0i}_{{\rm can}(1)} \right)_{\rm sym} =&& {4 \over c^3}
    \sum_{l \geq 1} {(-)^l \over l!}
    \partial_{L-1} \left\{{ M^{(1)}_{iL-1} (t-r/c) +
    M^{(1)}_{iL-1} (t+r/c) \over 2r } \right\} \nonumber \\
    && +\, {4 \over c^3} \sum_{l \geq 1} {(-)^l l \over (l+1)!}
    \varepsilon_{iab} \partial_{aL-1} \left\{ { S_{bL-1} (t-r/c) +
    S_{bL-1} (t+r/c) \over 2r} \right\}\ , \label{eq:2.5b} \\
 \left( h^{ij}_{{\rm can} (1)} \right)_{\rm sym} =&& - {4 \over c^4}
    \sum_{l \geq 2} {(-)^l \over l!}
    \partial_{L-2} \left\{{ M^{(2)}_{ijL-2} (t-r/c) +
    M^{(2)}_{ijL-2} (t+r/c) \over 2r } \right\} \nonumber \\
    && - {8 \over c^4} \sum_{l \geq 2} {(-)^l l \over (l+1)!}
  \partial_{aL-2} \left\{ \varepsilon_{ab(i}
{S^{(1)}_{j)bL-2}(t-r/c) + S^{(1)}_{j)bL-2} (t+r/c) \over 2r} \right\}\ .
    \label{eq:2.5c}
\end{eqnarray}
\end{mathletters}
The antisymmetric part is given similarly. However, as show in paper I, it
can be re-written profitably in the equivalent form
\begin{equation}
 \left( h^{\mu\nu}_{{\rm can} (1)} \right)_{\rm antisym} =
  - {4\over Gc^{2+s}} V^{\mu\nu}_{\rm reac} -\partial^\mu
  \xi^\nu -\partial^\nu \xi^\mu +\eta^{\mu\nu}\partial_\lambda \xi^\lambda
 \ . \label{eq:2.6}
\end{equation}
The second, third and fourth terms clearly represent a linear gauge
transformation, associated with the gauge vector $\xi^\mu$.
This vector is made of antisymmetric waves, and reads
\FL
\begin{mathletters}
\label{eq:2.7}
\begin{eqnarray}
 \xi^0  =&& {2 \over c} \sum_{l \geq 2}
   {(-)^l \over l!} { 2 l +1 \over l(l-1)} \partial_L
  \left\{{ M^{(-1)}_L (t-r/c) - M^{(-1)}_L (t+r/c) \over 2r } \right\}
  \ , \label{eq:2.7a} \\
 \xi^i  =&& - 2  \sum_{l \geq 2} {(-)^l \over l!}
    {(2l+1) (2l+3) \over l(l-1)} \partial_{iL}
  \left\{{ M^{(-2)}_L (t-r/c) - M^{(-2)}_L (t+r/c) \over 2r } \right\}
         \nonumber \\
    && + {4 \over c^2} \sum_{l \geq 2} {(-)^l \over l!}
    {2l+1 \over l-1} \partial_{L-1}
\left\{{ M_{iL-1}(t-r/c) - M_{iL-1}(t+r/c) \over 2r }\right\}\nonumber \\
    && +\, {4 \over c^2} \sum_{l \geq 2} {(-)^ll \over (l+1)!}
   {2l+1 \over l-1} \varepsilon_{iab} \partial_{aL-1}
 \left\{{ S^{(-1)}_{bL-1} (t-r/c) -  S^{(-1)} _{bL-1} (t+r/c) \over 2r }
     \right\} \ .  \label{eq:2.7b}
\end{eqnarray}
\end{mathletters}
Note that even though we have introduced first and second time-antiderivatives
of the multipole moments, denoted e.g. by
$M_{L}^{(-1)}(t)$ and $M_{L}^{(-2)}(t)$,
the dependence of (\ref{eq:2.7}) on the multipole moments ranges in fact
only in the time interval between $t-r/c$ and $t+r/c$ (see paper~I).
The first term in (\ref{eq:2.6}) defines, for our purpose, a
radiation reaction tensor potential $V^{\mu\nu}_{\rm reac}$ in the
linearized theory (in this term $s$ takes the values $0,1,2$ according
to $\mu\nu=00,0i,ij$).  The components of this potential
are given by \cite{N}
\FL
\begin{mathletters}
\label{eq:2.8}
\begin{eqnarray}
 V^{00}_{\rm reac} &&=  G
  \sum_{l \geq 2} {(-)^l \over l!} {(l+1) (l+2) \over l(l-1)}
   \hat\partial_L \left\{{ M_L (t-r/c) - M_L(t+r/c) \over 2r }\right\}
   \ , \label{eq:2.8a} \\
 V^{0i}_{\rm reac} &&=  -c^2 G
   \sum_{l \geq 2} {(-)^l \over l!} {(l+2) (2l+1) \over l(l-1)}
   \hat \partial_{iL} \left\{{     M^{(-1)} _L (t-r/c) -
    M^{(-1)} _L (t+r/c) \over 2r } \right\}  \nonumber \\
  && + G \sum_{l \geq 2} {(-)^l l \over (l+1)!}
    {l+2 \over l-1} \varepsilon_{iab} \hat\partial_{aL-1}
   \left\{ { S_{bL-1} (t-r/c) - S_{bL-1} (t+r/c) \over 2r} \right\}\ ,
      \label{eq:2.8b} \\
 V^{ij}_{\rm reac} &&=  c^4 G
    \sum_{l \geq 2} {(-)^l \over l!} {(2l+1) (2l+3) \over l(l-1)}
    \hat \partial_{ijL} \left\{{   M^{(-2)}_L(t-r/c) -
    M^{(-2)} _L (t+r/c) \over 2r } \right\} \nonumber \\
  && -2c^2 G  \sum_{l \geq 2} {(-)^l l \over (l+1)!} {2l+1 \over l-1}
    \varepsilon_{ab(i} \hat \partial_{j)aL-1} \left\{ {
   S^{(-1)}_{bL-1} (t-r/c) - S^{(-1)}_{bL-1} (t+r/c) \over 2r } \right\}
  \   \label{eq:2.8c}
\end{eqnarray}
\end{mathletters}
(see Eqs.  (2.19) of paper I).
By adding the contributions of the
gauge terms associated with (\ref{eq:2.7}) to the radiation reaction potential
(\ref{eq:2.8}) one reconstructs precisely, as stated by (\ref{eq:2.6}), the
antisymmetric part $(h^{\mu\nu}_{\rm can (1)})_{\rm antisym}$ of the
linearized field (\ref{eq:2.3}).

The scalar, vector, and tensor components of (\ref{eq:2.8}) generalize,
within the linearized theory, the Burke-Thorne \cite{Bu69,Th69,Bu71}
scalar potential by taking into account all multipolarities of waves,
and, in principle, all orders in the post-Newtonian expansion.  Actually
a full justification of this assertion would necessitate a matching to
the field inside the source, such as the one we perform in this paper at
1PN order. At the ``Newtonian'' order, the $00$ component of the potential
reduces to the Burke-Thorne potential,
\begin{equation}
 V^{00}_{\rm reac} = - {G\over 5c^5} x^i x^j M^{(5)}_{ij} (t)
    + O \left( {1\over c^7} \right)\ . \label{eq:2.9}
\end{equation}
At this order the $0i$ and $ij$ components make negligible
contributions.  Recall that a well-known property of the Burke-Thorne
reactive potential is to yield an energy loss in agreement with the
Einstein quadrupole formula, even though it is derived, in this
particular coordinate system, within the linearized theory (see
\cite{WalW80}).  In this paper we shall show that the same property
remains essentially true at the 1PN order. [This property is in general
false for other reactive potentials, valid in other coordinate systems, for
which the non-linear contributions play an important role.]
Evaluating the reaction potential $V^{\mu\nu}_{\rm reac}$ at the 1PN
order beyond (\ref{eq:2.9}), we find that both the
$00$ and $0i$ components are to be considered, and are
given by
\FL
\begin{mathletters}
\label{eq:2.10}
\begin{eqnarray}
 V^{00}_{\rm reac} = && - {G\over 5c^5} x^a x^b M^{(5)}_{ab} (t) +
  {G\over c^7} \left[ {1\over 189} x^a x^b x^c M^{(7)}_{abc} (t)
  - {1\over 70} r^2 x^a x^b M^{(7)}_{ab} (t)
   \right] + O\left( {1\over c^9} \right)\ , \label{eq:2.10a} \\
 V^{0i}_{\rm reac} = && {G\over c^5} \left[ {1\over 21} \hat x^{iab}
 M^{(6)}_{ab} (t)-{4\over 45} \varepsilon_{iab} x^a x^c S^{(5)}_{bc} (t)
  \right] + O\left( {1\over c^7} \right) \ . \label{eq:2.10b}
\end{eqnarray}
\end{mathletters}
At this order the $ij$ components of the potential can be neglected.

In the next subsection we address the question of the corrections to the
1PN reaction potential (\ref{eq:2.10}) which arise from the
non-linear contributions to the exterior field.  Answering this question
means controlling the non-linear metric at the 3.5PN order.

\subsection{The 3.5PN approximation in the exterior metric}

The radiation reaction potential (\ref{eq:2.8})-(\ref{eq:2.10}) represents
the antisymmetric part of the linearized metric in a particular coordinate
system, obtained
from the initial harmonic coordinate system in which
(\ref{eq:2.3}) holds by applying the gauge transformation associated
with (\ref{eq:2.7}).  In this coordinate system, the new
linearized metric reads
\begin{equation}
  h^{\mu\nu}_{(1)} = h^{\mu\nu}_{\rm can(1)} + \partial^\mu \xi^\nu
 + \partial^\nu \xi^\mu - \eta^{\mu\nu} \partial_\lambda \xi^\lambda\ .
  \label{eq:2.11}
\end{equation}
It fulfills, of course, the linearized equations (\ref{eq:2.2}).
Furthermore, since $\Box \xi^\mu =0$, the harmonic coordinate condition
is still satisfied.  However, we recall from paper I that the latter
new harmonic coordinate system is not completely satisfying for general
purposes, and should be replaced by a certain modified (non-harmonic)
coordinate system.  The reason is that the gauge vector defined by
(\ref{eq:2.7}) is made of antisymmetric waves, and consequently the
metric (\ref{eq:2.11}) contains both retarded {\it and} advanced waves.
In particular, the metric is no longer stationary in the remote past
(before the instant $-{\cal T}$ where the moments are assumed to be
constant).  Of course, the advanced waves which have been introduced are
pure gauge.  Nevertheless, the non-stationarity of the linearized metric
in the remote past breaks one of our initial assumptions, and this can
be a source of problems when performing the non-linear iterations of the
metric by this method.  Therefore, it was found necessary in paper I to
replace the gauge vector (\ref{eq:2.7}) by a modified gauge vector, such
that the modified coordinate system has two properties.  First, it
reduces in the near-zone of the source (the domain $D_i$ defined in
Sect.~III), to the un-modified coordinate system given by (\ref{eq:2.7}),
with a given but {\it arbitrary} post-Newtonian precision.  Second, it
reduces {\it exactly} to the initial harmonic coordinate system in which the
linearized metric is (\ref{eq:2.3}) in a domain exterior to some {\it
time-like} world tube surrounding the source (in fact, a future-oriented
time-like cone whose vertex is at the event $t={\cal -T}$, ${\bf x}={\bf
0}$).  This modified coordinate system is non-harmonic.  It has been
suggested to the author by T.  Damour in a private communication, and is
defined in Sect.  II.C of paper I.  By the first property, we see that
if we are interested to the field in the near-zone of the source, and
work at some finite post-Newtonian order (like the 1PN order
investigated in this paper), we can make all computations using the
un-modified gauge vector (\ref{eq:2.7}).  Indeed, it suffices to adjust
a certain constant, denoted $K$ in paper I, so that the modified gauge
vector agrees with (\ref{eq:2.7}) with a higher post-Newtonian
precision.  Thus, in the present paper, we shall not use explicitly the
modified coordinate system.  By the second property, we see that the
standard fall-off behavior of the metric at the various infinities
(notably the standard no-incoming radiation condition at past-null
infinity) are preserved in the modified coordinate system. By
these two properties, one can argue that what only matters is the {\it
existence} of such a modified coordinate system, which permits us to
make the connection with the field at infinity, but that in all
practical computations of the metric in the near-zone one can use the
un-modified coordinate system defined by (2.7).

Based on the linearized metric (\ref{eq:2.11}) (or, rather, on the modified
linearized metric (2.29) of paper~I) we built a full non-linear expansion,
\begin{equation}
  h^{\mu\nu} = G h^{\mu\nu}_{(1)} + G^2 h^{\mu\nu}_{(2)}
 + G^3 h^{\mu\nu}_{(3)} + ... \ ,  \label{eq:2.12}
\end{equation}
satisfying the vacuum field equations in a perturbative sense
(equating term by term the coefficients of equal powers of $G$ in both
sides of the equations). The non-linear coefficients
$h_{(2)}^{\mu\nu}$, $h_{(3)}^{\mu\nu}$, ...  are like
$h_{(1)}^{\mu\nu}$ in the form of multipole expansions parametrized by
$M_L$ and $S_L$. The construction of the non-linear metric is based on
the method of \cite{BD86}. As we are considering multipole expansions
valid in $D_e$ (and singular at the spatial origin $r=0$), we need to
use at each non-linear iterations of the field equations a special
operator generalizing the usual retarded integral operator when acting
on multipole expansions.  We denote this operator by ${\cal
F}\Box^{-1}_R$, to mean the ``Finite part of the retarded integral
operator'' (see \cite{BD86} for its precise definition).  The
non-linear coefficients $h_{(2)}^{\mu\nu}$, $h_{(3)}^{\mu\nu}$, ...
are given by
\begin{mathletters}
\label{eq:2.13}
\begin{eqnarray}
 h^{\mu\nu}_{(2)} &=& {\cal F}\ \Box^{-1}_R \Lambda_{(2)}^{\mu\nu} (h_{(1)})
    + q^{\mu\nu}_{(2)}\ , \label{eq:2.13a}  \\
 h^{\mu\nu}_{(3)} &=& {\cal F}\ \Box^{-1}_R \Lambda_{(3)}^{\mu\nu}
      (h_{(1)}, h_{(2)})
    + q^{\mu\nu}_{(3)}\ , \quad ...\ ,  \label{eq:2.13b}
\end{eqnarray}
\end{mathletters}
where the non-linear source terms $\Lambda^{\mu\nu}_{(2)}$,
$\Lambda^{\mu\nu}_{(3)}$,~...  represent the field non-linearities in
vacuum, and depend, at each non-linear order, on the
coefficients of the previous orders. The second
terms $q^{\mu\nu}_{(2)}$, $q^{\mu\nu}_{(3)}$,~...  ensure the
satisfaction of the harmonic coordinate condition at each non-linear
order (see \cite{BD86}).

When investigating the 3.5PN approximation, we can disregard purely
non-linear effects, such as the tail effect, which give irreducibly
non-local contributions in the metric inside the source.  These
effects arise at the 4PN approximation (see \cite{BD88} and Sect.~IV
below). Still there are some non-linear contributions in the metric at
the 3.5PN approximation, which are contained in the first two
non-linear coefficients $h_{(2)}^{\mu\nu}$ and $h_{(3)}^{\mu\nu}$
given by (\ref{eq:2.13}).  These contributions involve some non-local
integrals, but which ultimately do not enter the inner metric (after
matching).  As shown in paper I, the contributions due to
$h^{\mu\nu}_{(2)}$ and $h^{\mu\nu}_{(3)}$ in the 1PN radiation
reaction potential imply only a modification of the multipole moments
$M_L$ and $S_L$ parametrizing the potential.  We define two new sets
of multipole moments,
\begin{mathletters}
\label{eq:2.14}
\begin{eqnarray}
 \widetilde M_L (t) =&& M_L(t) + \left\{
 \begin{array}{c}
{G\over c^7} m(t) \  {\rm for} \, l=0 \\  \\
{G\over c^5} m_i (t) \ {\rm for} \, l=1 \\ \\
 0  \qquad {\rm for}\ l\geq 2
 \end{array} \right\}
   + {G\over c^7} T_L (t) + O \left( {1\over c^8} \right)\ ,
\label{eq:2.14a} \\
 \widetilde S_L (t) =&& S_L(t) + \left\{
 \begin{array}{c}
{G\over c^5} s_i (t) \ {\rm for} \, l=1 \\  \\
 0  \quad {\rm for}\ l\geq 2
\end{array} \right\}
    + O \left( {1\over c^6} \right) \ , \label{eq:2.14b}
\end{eqnarray}
\end{mathletters}
where the functions $m$, $m_i$ and $s_i$ are given by the
non-local expressions
\FL
\begin{mathletters}
\label{eq:2.15}
\begin{eqnarray}
  m (t) =&& - {1 \over 5}
    \int^t_{- \infty} dv\ M^{(3)}_{ab} (v) M^{(3)}_{ab} (v) +
      F (t) \ , \label{eq:2.15a}\\
 m_i (t)  =&& -{2\over 5} M_a M^{(3)}_{ia} (t)
 - {2\over 21c^2} \int^t_{- \infty} dv M^{(3)}_{iab} (v) M^{(3)}_{ab} (v)
    \nonumber \\
  && + {1\over c^2} \int^t_{-\infty} dv \int^v_{-\infty} dw
    \left[ - {2 \over 63} M^{(4)}_{iab} (w) M^{(3)}_{ab} (w)
    - {16 \over 45} \varepsilon_{iab} M^{(3)}_{ac} (w) S^{(3)}_{bc} (w)
       \right] + {1\over c^2} G_i (t) \ ,  \nonumber \\
   && \label{eq:2.15b} \\
  s_i (t) =&& - {2 \over 5} \varepsilon_{iab}
   \int^t_{- \infty} dv\ M^{(2)}_{ac} (v) M^{(3)}_{bc} (v) + H_i (t) \ .
    \label{eq:2.15c}
\end{eqnarray}
\end{mathletters}
The function $T_L (t)$ in (\ref{eq:2.14a}), and the functions $F(t)$,
$G_i(t)$ and $H_i(t)$ in (\ref{eq:2.15}), are some local (or
instantaneous) functions, which do not play a very important role
physically (they are computed in paper~I).  Then the radiation
reaction potential at the 1PN order, in the non-linear theory, is
given by the same expression as in (\ref{eq:2.10}), but expressed in
terms of the new multipole moments, say $\widetilde{\cal M}
\equiv \{\widetilde{M}_L, \widetilde{S}_L\}$,
\begin{mathletters}
\label{eq:2.16}
\begin{eqnarray}
 V^{\rm reac} [\widetilde {\cal M}] =&& - {G\over 5c^5} x^a x^b
 \widetilde M^{(5)}_{ab} (t) + {G\over c^7} \left[ {1\over 189} x^{abc}
 \widetilde M^{(7)}_{abc}(t) - {1\over 70} r^2 x^{ab} \widetilde M^{(7)}_{ab}
  (t) \right]  \nonumber \\
 && + O \left( {1\over c^8} \right) \ ,\label{eq:2.16a} \\
 V_i^{\rm reac} [\widetilde {\cal M}] =&&  {G\over c^5}
 \left[ {1\over 21} \hat x^{iab}  \widetilde M^{(6)}_{ab} (t)
  - {4\over 45}\varepsilon_{iab} x^{ac} \widetilde S^{(5)}_{bc} (t)
   \right] + O \left( {1\over c^6} \right) \  \label{eq:2.16b}
\end{eqnarray}
\end{mathletters}
(see Eq. (3.53) in paper~I).  In the considered coordinate system, the
metric, accurate to 1PN order as concerns both the usual non-radiative
effects and the radiation reaction effects, reads
(coming back to the usual covariant metric $g^{\rm ext}_{\mu\nu}$)
\FL
\begin{mathletters}
\label{eq:2.17}
\begin{eqnarray}
 g^{\rm ext}_{00} =&& - 1 + {2\over c^2}
\left( V^{\rm ext} [\widetilde{\cal M}] + V^{\rm reac}[\widetilde{\cal M}] 
\right)
  - {2\over c^4} \left( V^{\rm ext} [\widetilde{\cal M}]
 + V^{\rm reac} [\widetilde{\cal M}] \right)^2 \nonumber \\
&&+{1\over c^6}\,{}_6 g^{\rm ext}_{00}+{1\over c^8}\,{}_8 g^{\rm ext}_{00}
    + O \left( {1\over c^{10}} \right)\ ,   \label{eq:2.17a} \\
 g^{\rm ext}_{0i} =&&  - {4\over c^3}
  \left( V^{\rm ext}_i[\widetilde{\cal M}]+V_i^{\rm reac}
  [\widetilde{\cal M}]\right)
  + {1\over c^5}\,{}_5 g^{\rm ext}_{0i}+{1\over c^7}\,{}_7 g^{\rm ext}_{0i}
    + O \left( {1\over c^9} \right)\ ,  \label{eq:2.17b} \\
 g^{\rm ext}_{ij} =&& \delta_{ij} \left[ 1 + {2\over c^2}
\left( V^{\rm ext}[\widetilde{\cal M}]+V^{\rm reac} [\widetilde{\cal M}] 
\right)
\right] + {1\over c^4} {}_4g^{\rm ext}_{ij} + {1\over c^6} {}_6g^{\rm ext}_{ij}
    + O \left( {1\over c^8} \right)\ .  \label{eq:2.17c}
\end{eqnarray}
\end{mathletters}
The superscript ext is to remember that the metric is valid in the exterior
domain $D_e$, and will differ from the inner metric by a coordinate
transformation (see Sect.~III). The Newtonian and 1PN approximations are
entirely contained in the externals potentials $V^{\rm ext}$ and $V^{\rm
ext}_i$, given by the multipole expansions of {\it symmetric} waves,
\begin{mathletters}
\label{eq:2.18}
\begin{eqnarray}
 V^{\rm ext} [\widetilde{\cal M}] = &&G \sum_{l\geq 0} {(-)^l\over l!}
 \partial_L
 \left\{ {\widetilde M_L \left( t-{r\over c}\right) + \widetilde M_L
 \left( t+{r\over c}\right) \over 2r} \right\}\ , \label{eq:2.18a} \\
 V^{\rm ext}_i [\widetilde {\cal M}] =&& -G \sum_{l\geq 1} {(-)^l\over l!}
  \partial_{L-1}\left\{ {\widetilde M^{(1)}_{iL-1} \left( t-{r\over c}\right)
   + \widetilde M^{(1)}_{iL-1} \left( t+{r\over c}\right) \over 2r} \right\}
      \nonumber \\
  && -G \sum_{l\geq 1} {(-)^ll\over (l+1)!} \varepsilon_{iab}
  \partial_{aL-1}   \left\{ {\widetilde S_{bL-1} \left( t-{r\over c}\right)
 + \widetilde S_{bL-1}\left( t+{r\over c}\right)\over 2r}\right\}\ .
    \label{eq:2.18b}
\end{eqnarray}
\end{mathletters}
The 2PN and 3PN approximations are not controlled at this stage; they
are symbolized in (\ref{eq:2.17}) by the terms $c^{-n} {}_ng^{\rm
ext}_{\mu\nu}$.  However, these approximations are non-radiative (or
non-dissipative), as are the Newtonian and 1PN approximations (the
1PN, 2PN, and 3PN terms are ``even" in the sense that they yield only
even powers of $1/c$ in the equations of motion). A discussion of the
4PN approximation in the exterior metric can be found in Sect.~III.D
of paper I.

As it stands, the metric (\ref{eq:2.17}) is disconnected from the actual
source of radiation. The multipole moments
$\widetilde M_L$ and $\widetilde S_L$ are left as some un-specified
functions of time.  Therefore, in order to determine the radiation
reaction potentials (\ref{eq:2.16}) as some {\it explicit} functionals
of the matter variables, we need to relate the exterior metric
(\ref{eq:2.17}) to a metric valid inside the source, solution of the
non-vacuum Einstein field equations.  We perform the relevant
computation in the next section, and obtain the multipole moments
$\widetilde M_L$ and $\widetilde S_L$ as integrals over the source.

\section{The 1PN-accurate radiation reaction potentials}  \label{sec:3}

\subsection{The inner gravitational field}  

The near-zone of the source is defined in the usual way as being
an inner domain $D_i= \{({\bf x},t), |{\bf x}|<r_i\}$, whose radius
$r_i$ satisfies $r_i>a$ ($D_i$ covers entirely the source), and
$r_i\ll\lambda$ (the domain $D_i$ is of small extent as compared with one
wavelength of the radiation).  These two demands are possible
simultaneously when the source is slowly moving, i.e.  when there exists
a small parameter of the order of $1/c$ when $c\to \infty$, in
which case we can assume $r_i/\lambda =O(1/c)$.  Furthermore, we can
adjust $r_e$ and $r_i$ so that $a<r_e<r_i$ (where $r_e$
defines the external domain $D_e$ in Sect.~II).

In this subsection we present the result of the expression of the
metric in $D_i$, to the relevant approximation. In the next subsection
we prove that this metric matches to the exterior metric reviewed in
Sect.~II.  The accuracy of the inner metric is 1PN for the usual
non-radiative approximations, and 1PN for the dominant radiation
reaction. Thus, the metric enables one to control, in the equations of
motion, the Newtonian acceleration followed by the first relativistic
correction, which is of order $c^{-2}$ or 1PN, then the dominant
``Newtonian'' radiation reaction, of order $c^{-5}$ or 2.5PN, and
finally the first relativistic 1PN correction in the reaction,
$c^{-7}$ or 3.5PN.  The intermediate approximations $c^{-4}$ and
$c^{-6}$ (2PN and 3PN) are left un-determined, like in the exterior
metric (2.17).  The inner metric, in $D_i$, reads
\begin{mathletters}
\label{eq:3.1}
\begin{eqnarray}
 g^{\rm in}_{00} &=& -1 +{2\over c^2} {\cal V}^{\rm in} - {2\over c^4} 
  ({\cal V}^{\rm in})^2 +
  {1\over c^6}{}_6 g^{\rm in}_{00} + {1\over c^8} {}_8 g^{\rm in}_{00}
  + O\left({1\over c^{10}}\right)\ , \label{eq:3.1a} \\
 g^{\rm in}_{0i} &=& -{4\over c^3} {\cal V}^{\rm in}_i + {1\over c^5} {}_5
 g^{\rm in}_{0i} + {1\over c^7} {}_7 g^{\rm in}_{0i}+ O\left({1\over
    c^9}\right)\ ,\label{eq:3.1b} \\
 g^{\rm in}_{ij} &=& \delta_{ij} \left(1 +{2\over c^2}{\cal V}^{\rm in}\right)
    + {1\over c^4} {}_4 g^{\rm in}_{ij} + {1\over c^6} {}_6 g^{\rm in}_{ij}
    + O\left({1\over c^8}\right) \ .  \label{eq:3.1c}
\end{eqnarray}
\end{mathletters}
It is valid in a particular Cartesian coordinate system $({\bf x},t)$,
which is to be determined by matching. Like in (2.17), the terms
$c^{-n} {}_n g^{\rm in}_{\mu\nu}$ represent the 2PN and 3PN
approximations.  Note that these terms depend functionally on the
source's variables through some {\it spatial} integrals, extending
over the whole three-dimensional space, but that they do not involve
any non-local integral in time.  These terms are ``instantaneous'' (in
the terminology of \cite{BD88}) and ``even'', so they remain invariant
in a time reversal, and do not yield any radiation reaction
effects. The remainder terms in (\ref{eq:3.1}) represent the 4PN and
higher approximations.  Note also that some logarithms of $c$ arise
starting at the 4PN approximation.  For simplicity we do not indicate
in the remainders the dependence on $\ln c$.  The potentials ${\cal
V}^{\rm in}$ and ${\cal V}^{\rm in}_i$ introduced in (\ref{eq:3.1})
are given, like in (\ref{eq:2.17}), as the linear combination of two
types of potentials. With the notation ${\cal V}^{\rm in}_\mu \equiv
({\cal V}^{\rm in}, {\cal V}^{\rm in}_i)$, where the index $\mu$ takes
the values $0,i$ and where ${\cal V}^{\rm in}_0\equiv {\cal V}^{\rm
in}$, we have
\begin{equation}
 {\cal V}^{\rm in}_\mu = V^{\rm in}_\mu [\sigma_\nu] + V^{\rm reac}_\mu
   [{\cal I}]\ .  \label{eq:3.2}
\end{equation}

The first type of potential, $V^{\rm in}_\mu$, is given by an integral
of the {\it symmetric} potentials, i.e.  by the half-sum of the
retarded integral and of the corresponding advanced integral.  Our
terminology, which means here something slightly different from
Sect.~II, should be clear from the context.  We are referring here to
the formal structure of the integral, made of the sum of the retarded
and advanced integrals. However, the real behavior of the symmetric
integral under a time-reversal operation may be more complicated than
a simple invariance.  The mass and current densities $\sigma_\mu
\equiv (\sigma ,
\sigma_i)$ of the source are defined by
\begin{mathletters}
\label{eq:3.3}
\begin{eqnarray}
  \sigma &\equiv & {T^{00} + T^{kk}\over c^2}\ , \label{eq:3.3a}\\
  \sigma_i &\equiv & {T^{0i}\over c}\ , \label{eq:3.3b}
\end{eqnarray}
\end{mathletters}
where $T^{\mu\nu}$ denotes the usual stress-energy tensor of the
matter fields (with $T^{kk}$ the spatial trace $\Sigma \delta_{jk}
T^{jk}$). The powers of $1/c$ in (\ref{eq:3.3}) are such that
$\sigma_\mu$ admits a finite non-zero limit when $c\to +\infty$.  The
potentials $V^{\rm in}_\mu$ are given by
\begin{equation}
  V^{\rm in}_\mu ({\bf x},t) = {G\over 2} \int {d^3{\bf x}'\over |{\bf
   x}-{\bf x}'|} \left[ \sigma_\mu \left({\bf x}',t -{1\over c}
    |{\bf x}-{\bf x}'|\right) + \sigma_\mu \left({\bf x}',t +{1\over c}
    |{\bf x}-{\bf x}'|\right) \right]\ . \label{eq:3.4}
\end{equation}
To lowest order when $c\to +\infty$, $V^{\rm in}$ reduces to the usual
Newtonian potential, and $V^{\rm in}_i$ to the usual gravitomagnetic
potential. It was noticed in \cite{BD89} that when using the mass
density $\sigma$ given by (3.3a), the first (non-radiative)
post-Newtonian approximation takes a very simple form, involving
simply the square of the potential in the $00$ component of the
metric.  See also (4.5) below, where we use the post-Newtonian
expansion $V^{\rm in} = U+ \partial^2_t X/2c^2 + O(c^{-4})$.

The fact that the inner metric contains some symmetric integrals, and
therefore some advanced integrals, does not mean that the field
violates the condition of retarded potentials. The metric
(\ref{eq:3.1}) is in the form of a post-Newtonian expansion, which is
valid only in the near zone $D_i$.  It is well-known that the
coefficients of the powers of $1/c$ in a post-Newtonian expansion
typically diverge at spatial infinity. This is no concern of us
because the expansion is not valid at infinity (it would give poor
results when compared to an exact solution).  Thus, the symmetric
integral (\ref{eq:3.4}) should more properly be replaced by its formal
post-Newtonian expansion, readily obtained by expanding by means of
Taylor's formula the retarded and advanced arguments when $c \to
+\infty$.  Denoting $\partial^{2p}_t \equiv (\partial/\partial
t)^{2p}$ we have
\begin{equation}
  V^{\rm in}_\mu ({\bf x},t) = G \sum^{+\infty}_{p=0} {1\over (2p)!c^{2p}}
 \int d^3 {\bf x}' |{\bf x}-{\bf x}'|^{2p-1} \partial^{2p}_t \sigma_\mu
  ({\bf x}',t)\ . \label{eq:3.5}
\end{equation}
This expansion could be limited to the precision indicated in
(\ref{eq:3.1}). It involves (explicitly) only {\it even} powers of
$c^{-1}$, and is thus expected to yield essentially non-dissipative
effects.  However, the dependence of $V^{\rm in}_\mu$ on $c^{-1}$ is
more complicated than indicated in (\ref{eq:3.5}).  Indeed, the mass
and current densities $\sigma_{\mu}$ depend on the metric
(\ref{eq:3.1}), and thus depend on $c^{-1}$ starting at the
post-Newtonian level.  Even more, the densities $\sigma_{\mu}$ and
their time-derivatives do contain, through the contribution of the
reactive potentials $V^{\rm reac}_\mu$ (see below), some odd powers of
$c^{-1}$ which are associated ({\it a priori}) to radiation reaction
effects.  These ``odd'' contributions in $V^{\rm in}_\mu$ form an
integral part of the equations of motion of binary systems at the
3.5PN approximation \cite{IW95}, but we shall prove in Sect.~IV that
they do not participate to the losses of energy and momenta by
radiation at the 1PN order, as they enter the balance equations only
in the form of some total time-derivatives.  The secular losses of
energy and momenta are driven by the radiation reaction potentials
$V^{\rm reac}_\mu$, to which we now turn.

The 1PN-accurate reaction potentials $V^{\rm reac}_\mu \equiv (V^{\rm
reac}, V^{\rm reac}_i)$ involve dominantly some odd powers of
$c^{-1}$, which correspond in the metric (\ref{eq:3.1}) to the 2.5PN
and 3.5PN approximations $c^{-5}$ and $c^{-7}$ taking place between
the (non-dissipative) 2PN and 3PN approximations. Since the reactive
potentials are added linearly to the potentials $V^{\rm in}_\mu$, the
simple form mentionned above of the 1PN non-radiative approximation
holds also for the 1PN radiative approximation, in this coordinate
system. The $V^{\rm reac}_\mu$'s are given by exactly the same
expressions as obtained in paper~I for the exterior metric (see
(\ref{eq:2.16}) in Sect.~II), but they depend on some specific
``source'' multipole moments ${\cal I}\equiv \{I_L,J_L\}$ instead of
the unknown multipole moments $\widetilde{\cal M}$.  Namely,
\begin{mathletters}
\label{eq:3.6}
\begin{eqnarray}
V^{\rm reac}({\bf x},t)&=& -{G\over 5c^5} x_{ij} I^{(5)}_{ij} (t) +
  {G\over c^7} \left[ {1\over 189} x_{ijk} I^{(7)}_{ijk} (t) - {1\over
  70} {\bf x}^2 x_{ij} I^{(7)}_{ij} (t) \right] + O \left( {1\over
  c^8} \right) \ , \label{eq:3.6a}\\ V_i^{\rm reac}({\bf x},t) &=&
  {G\over c^5} \left[ {1\over 21} \hat x_{ijk} I^{(6)}_{jk} (t) -
  {4\over 45} \varepsilon_{ijk} x_{jm} J^{(5)}_{km} (t) \right] + O
  \left( {1\over c^6} \right)\ ,\label{eq:3.6b}
\end{eqnarray}
\end{mathletters}
where we recall our notation $\hat x_{ijk} = x_{ijk} - {1\over 5} {\bf
x}^2 (\delta_{ij} x_k +\delta_{ik}x_j + \delta_{jk}x_i)$ \cite{N}. The
multipole moments $I_{ij}(t)$, $I_{ijk}(t)$ and $J_{ij}(t)$ are some
explicit functionals of the densities $\sigma_{\mu}$.  Only the mass
quadrupole $I_{ij}(t)$ in the first term of $V^{\rm reac}$ needs to be
given at 1PN order.  The relevant expression is
\begin{equation}
 I_{ij} = \int d^3{\bf x} \left\{ \hat x_{ij} \sigma + {1\over 14c^2}
 {\bf x}^2 \hat x_{ij} \partial_t^2 \sigma - {20\over 21c^2} \hat x_{ijk}
\partial_t \sigma_k  \right\}\ \label{eq:3.7}
\end{equation}
(see (\ref{eq:3.21a}) for the general expression of $I_L$).
 The mass octupole and current quadrupole $I_{ijk}(t)$
and $J_{ij}(t)$
take their standard Newtonian expressions,
\begin{mathletters}
 \label{eq:3.8}
\begin{eqnarray}
  I_{ijk} &=& \int d^3 {\bf x}\,\hat x_{ijk} \sigma + O \left( {1\over c^2}
    \right)\ ,     \label{eq:3.8a}\\
  J_{ij} &=& \int d^3 {\bf x}\, \varepsilon_{km<i} \hat x_{j>k} \sigma_m
    \ .   \label{eq:3.8b}
\end{eqnarray}
\end{mathletters}
The potentials $V^{\rm reac}$ and $V^{\rm reac}_i$ generalize to 1PN
order the scalar reactive potential of Burke and Thorne
\cite{Bu69,Th69,Bu71}, whose form is that of the first term in
(\ref{eq:3.6a}). The vectorial potential $V^{\rm reac}_i$ enters the
equations of motion at the same 3.5PN order as the 1PN corrections in
$V^{\rm reac}$.  The first term which is neglected in $V^{\rm reac}$,
of order $c^{-8}$ or 1.5PN, is due to the tails of waves (see
Sect.~IV.C).

\subsection{Matching to the exterior field}  

In this subsection we prove that the inner metric presented above (i)
satisfies the Einstein field equations within the source (in the
near-zone $D_i$), and (ii) matches to the exterior metric
(\ref{eq:2.17}) in the intersecting region between $D_i$ and the
exterior zone $D_e$ (exterior near-zone $D_i\cap D_e$).

Note that during Proof (i) we do not check any boundary conditions
satisfied by the metric at infinity.  Simply we prove that the metric
satisfies the field equations term by term in the post-Newtonian
expansion, but at this stage the metric could be made of a mixture of
retarded and advanced solutions.  Only during Proof (ii) does one
check that the metric comes from the re-expansion when $c \to \infty$
of a solution of the Einstein field equations satisfying some relevant
time-asymmetric boundary conditions at infinity.  Indeed, the exterior
metric has been constructed in paper~I by means of a post-Minkowskian
algorithm valid all over the exterior region $D_e$, and having a
no-incoming radiation condition built into it (indeed, the exterior
metric was assumed to be stationary in the remote past, see Sect.~II).

The proof that (\ref{eq:3.1}) is a solution admissible
in $D_i$ follows immediately from the particular form
taken by the Einstein field equations when
developed to 1PN order \cite{BD89}
\begin{mathletters}
\label{eq:3.9}
\begin{eqnarray}
 \Box \ln (-g^{\rm in}_{00}) &=& {8\pi G\over c^2} \sigma + 
  O_{\rm even}\left({1\over c^{6}}\right)\ , \label{eq:3.9a} \\
 \Box g^{\rm in}_{0i} &=& {16\pi G\over c^3} \sigma_i +
  O_{\rm even}\left({1\over c^{5}}\right)\ ,\label{eq:3.9b}\\
 \Box g^{\rm in}_{ij} &=& -{8\pi G\over c^2} \delta_{ij} \sigma +
  O_{\rm even}\left({1\over c^{4}}\right) \ .  \label{eq:3.9c}
\end{eqnarray}
\end{mathletters}
The point is that with the introduction of the logarithm of $-g^{\rm
in}_{00}$ as a new variable in (\ref{eq:3.9a}) the equations at the
1PN order take the form of {\it linear} wave equations.  The other
point is that the neglected post-Newtonian terms in (\ref{eq:3.9}) are
``even", in the sense that the {\it explicit} powers of $c^{-1}$ they
contain, which come from the differentiations of the metric with
respect to the time coordinate $x^0=c t$, correspond formally to
integer post-Newtonian approximations (to remember this we have added
the subscript ``even'' on the $O$-symbols).  This feature is simply
the consequence of the time-symmetry of the field equations, implying
that to each solution of the equations is associated another solution
obtained from it by a time-reversal.

Because the potentials $V^{\rm in}_\mu$ satisfy exactly $\Box V^{\rm
in}_\mu = - 4 \pi G \sigma_{\mu}$, a consistent solution of
(\ref{eq:3.9}) is easily seen to be given by (\ref{eq:3.1}) in which
the reactive potentials $V^{\rm reac}_\mu$ are set to zero.  Now the
equations (\ref{eq:3.9}) are linear wave equations, so we can add
linearly to $V^{\rm in}_\mu$ any homogeneous solution of the wave
equation which is regular in $D_i$.  One can check from their
definition (\ref{eq:3.6}) that the reactive potentials $V^{\rm
reac}_\mu$ form such a homogeneous solution, as they satisfy
\begin{mathletters}
\label{eq:3.10}
\begin{eqnarray}
\Box V^{\rm reac} = O\left({1\over c^{8}}\right), \label{eq:3.10a}\\
\Box V^{\rm reac}_i = O\left( {1\over c^{6}} \right) . \label{eq:3.10b}
\end{eqnarray}
\end{mathletters}
So we can add $V^{\rm reac}_\mu$ to $V^{\rm in}_\mu$, defining an
equally consistent solution of (\ref{eq:3.9}), which is precisely
(\ref{eq:3.1}), modulo the error terms coming from (\ref{eq:3.10}) and
which correspond to the neglected 4PN approximation. [Note that
$V^{\rm reac}_\mu$ comes from the expansion of the tensor potential
(\ref{eq:2.8}) of the linearized theory, which satisfies {\it exactly}
the source-free wave equation.  It would be possible to define $V^{\rm
reac}_\mu$ in such a way that there are no error terms in
(\ref{eq:3.10}). See (\ref{eq:4.33}) for more precise expressions of
$V^{\rm reac}_{\mu}$, satisfying more precisely the wave equation.]

With Proof (i) done, we undertake Proof (ii). More precisely, we show
that (\ref{eq:3.1}) differs from the exterior metric (\ref{eq:2.17})
by a mere coordinate transformation in $D_i\cap D_e$.  This will be
true only if the multipole moments $\widetilde M_L$ and $\widetilde
S_L$ parametrizing (\ref{eq:2.17}) agree, to the relevant order, with
some source multipole moments $I_L$ and $J_L$.  Fulfilling these
matching conditions will ensure (in this approximate framework) the
existence and consistency of a solution of the field equations valid
everywhere in $D_i$ and $D_e$.  As recalled in the introduction, this
is part of the method to work out first the exterior metric leaving
the multipole moments arbitrary (paper~I), and then to obtain by
matching the expressions of these moments as integrals over the source
(this paper).

To implement the matching we expand the inner metric (\ref{eq:3.1})
into multipole moments outside the compact support of the source.  The
comparison can then be made with (\ref{eq:2.17}), which is already in
the form of a multipole expansion.  Only the potentials $V^{\rm
in}_\mu$ need to be expanded into multipoles, as the reactive
potentials $V^{\rm reac}_\mu$ are already in the required form. The
multipole expansion of the retarded integral of a compact-supported
source is well-known.  E.g., the formula has been obtained in the
Appendix B of \cite{BD89} using the STF formalism for spherical
harmonics.  The multipole expansion corresponding to an advanced
integral follows simply from the replacement $c\to -c$ in the formula.
The script letter ${\cal M}$ will be used to denote the multipole
expansion.  ${\cal M} (V^{\rm in}_\mu)$ reads as
\begin{mathletters}
\label{eq:3.11}
\begin{eqnarray}
 {\cal M} (V^{\rm in}) &=& G \sum_{l\geq 0} {(-)^{l} \over l !}
\partial_L \left \{ {F_L (t-r/c) + F_L(t+r/c) \over 2r} \right \}\,
\label{eq:3.11a}\\
 {\cal M} (V_i^{\rm in}) &=& G \sum_{l\geq 0} {(-)^l \over l !}
 \partial_L \left\{ {G_{iL} (t-r/c) +G_{iL}(t+r/c)\over 2r} \right\}
  \ ,  \label{eq:3.11b}
\end{eqnarray}
\end{mathletters}
where $F_L(t)$ and $G_{iL}(t)$ are some tensorial functions of time
given by the integrals
\begin{mathletters}
\label{eq:3.12}
\begin{eqnarray}
 F_L (t) &=& \int d^3{\bf x}\, \hat x_L \int^1_{-1} dz\,\delta_l
  (z) \sigma ({\bf x},t +z|{\bf x}|/c)\ , \label{eq:3.12a}\\
 G_{iL} (t) &=& \int d^3{\bf x}\, \hat x_L \int^1_{-1} dz\,\delta_l
  (z) \sigma_i ({\bf x},t +z|{\bf x}|/c)\ . \label{eq:3.12b}
\end{eqnarray}
\end{mathletters}
The function $\delta_l (z)$ appearing here takes into account the delays
in the propagation of the waves inside the source. It reads
\begin{equation}
  \delta_l (z) = {(2l+1)!!\over 2^{l+1}l !} (1-z^2)^l\ ;
  \quad  \int^1_{-1} dz \delta_l (z) = 1 \  \label{eq:3.13}
\end{equation}
(see Eq.~(B.12) in \cite{BD89}). Note that the same functions $F_L(t)$
and $G_{iL}(t)$ parametrize both the retarded and the corresponding
advanced waves in (\ref{eq:3.11}). Indeed $\delta_l (z)$ is an even
function of its variable $z$, so the integrals (\ref{eq:3.12}) are
invariant under the replacement $c\to -c$.

Using an approach similar to the one employed in \cite{BD89}, we
perform an irreducible decomposition of the tensorial function
$G_{iL}$ (which is STF with respect to its $l$ indices $L$ but not
with respect to its $l+1$ indices $iL$), as a sum of STF tensors of
multipolarities $l+1$, $l$ and $l-1$.  The equation (2.17) in
\cite{BD89} gives this decomposition as
\begin{equation}
 G_{iL} = C_{iL}  - {l\over l+1}
\varepsilon_{ai<i_l} D_{L-1>a} + {2l-1\over 2l+1}
\delta_{i<i_l} E_{L-1>} \ , \label{eq:3.14}
\end{equation}
where the tensors $C_{L+1}$, $D_L$ and $E_{L-1}$ (which are STF with
respect to all their indices) are given by
\begin{mathletters}
\label{eq:3.15}
\begin{eqnarray}
 C_{L+1}(t) &=& \int d^3 {\bf x} \int^1_{-1} dz \delta_l (z)
\hat x_{<L} \sigma_{i_{l+1}>} ({\bf x},t+z|{\bf x}|/c)\ ,\label{eq:3.15a}\\
 D_L(t) &=& \int d^3 {\bf x} \int^1_{-1} dz \delta_l (z)
  \varepsilon_{ab<i_l} \hat x_{L-1>a} \sigma_b ({\bf x},t+z|{\bf
  x}|/c)\ , \label{eq:3.15b}\\ E_{L-1}(t) &=& \int d^3 {\bf x}
  \int^1_{-1} dz \delta_l (z) \hat x_{aL-1} \sigma_a ({\bf x},t+z|{\bf
  x}|/c)\ . \label{eq:3.15c}
\end{eqnarray}
\end{mathletters}
Then by introducing the new definitions of STF tensors,
\begin{mathletters}
\label{eq:3.16}
\begin{eqnarray}
  A_L &=& F_L - {4(2l+1)\over c^2 (l+1)(2l+3)} E^{(1)}_L
  \ ,  \label{eq:3.16a}\\
  B_L &=& l\, C_L - {l\over c^2 (l+1)(2l+3)} E^{(2)}_L \ ,
    \label{eq:3.16b}
\end{eqnarray}
\end{mathletters}
and by using standard manipulations on STF tensors, we can re-write
the multipole expansions (\ref{eq:3.11}) in the new form
\begin{mathletters}
\label{eq:3.17}
\begin{eqnarray}
 {\cal M} (V^{\rm in})  &=& -c \partial_t \phi^0+ G
  \sum_{l\geq 0} {(-)^l\over l !} \partial_L \left\{ {A_L(t-r/c)
  +A_L (t+r/c)\over 2r} \right\}\ , \label{eq:3.17a}\\
 {\cal M} (V^{\rm in}_i)  &=& {{c^3} \over 4} \partial_i \phi^0 -G
 \sum_{l\geq 1} {(-)^l\over l !} \partial_{L-1} \left\{ {B_{iL-1}
 (t-r/c) +B_{iL-1} (t+r/c)\over 2r} \right\} \nonumber \\
 && -G\sum_{l\geq 1} {(-)^l\over l !} {l\over l+1}
   \varepsilon_{iab}\partial_{aL-1} \left\{ {D_{bL-1}
   (t-r/c) +D_{bL-1} (t+r/c)\over 2r} \right\}\ , \label{eq:3.17b}
\end{eqnarray}
\end{mathletters}
where we denote
\begin{equation}
 \phi^0 = - {4G \over {c^3}} \sum_{l\geq 0} {(-)^l\over(l+1)!}
{2l+1\over 2l+3}
  \partial_L \left\{ {E_L(t-r/c) +E_L (t+r/c)\over 2r} \right\}\ .
  \label{eq:3.18}
\end{equation}
Next the moment $A_L$ is expanded when $c \to \infty$ to 1PN order, and
the moments $B_L$, $D_L$ to Newtonian order.  The required formula
is (B.14) in \cite{BD89}, which immediately gives
\begin{mathletters}
\label{eq:3.19}
\begin{eqnarray}
 A_L &=& \int d^3 {\bf x} \left\{ \hat x_L \sigma+ {1\over 2c^2(2l+3)}
  {\bf x}^2 \hat x_L \partial^2_t \sigma - {4(2l+1)\over c^2(l+1)
  (2l+3)} \hat x_{iL} \partial_t \sigma_i \right\} \nonumber\\
  &&\qquad\quad + O_{\rm even}\left({1\over c^4} \right)\ ,
  \label{eq:3.19a} \\ B_L &=& l \int d^3 {\bf x}\, \hat x_{<L-1}
  \sigma_{i_l >} + O_{\rm even}\left({1\over c^2}\right)\ ,
  \label{eq:3.19b} \\ D_L &=& \int d^3 {\bf x}\,\varepsilon_{ab<i_l}
  \hat x_{L-1>a} \sigma_b + O_{\rm even}\left({1\over c^2} \right) \
  . \label{eq:3.19c}
\end{eqnarray}
\end{mathletters}
Here the notation $O_{\rm even}(c^{-n})$ for the post-Newtonian
remainders simply indicates that the whole post-Newtonian expansion is
composed only of even powers of $c^{-1}$, like in (\ref{eq:3.5}) (the
source densities $\sigma_\mu$ being considered to be independent of
$c^{-1}$), as clear from (B.14) in \cite{BD89}.  We now transform the
leading-order term in the equation for $B_L$ using the equation of
continuity for the mass density $\sigma$.  The Newtonian equation of
continuity does the needed transformation, but one must be careful
about the higher-order post-Newtonian corrections which involve some
reactive contributions.  It can be checked that these reactive
contributions arise only at the order $O(c^{-7})$, so that the
equation of continuity reads, with evident notation, $\partial_t\sigma
+\partial_i\sigma_i = O_{\rm even} (c^{-2}) +O(c^{-7})$.  From this
one deduces
\begin{equation}
 B_L = {d \over {dt}} \left\{ \int d^3 {\bf x} \hat x_L \sigma
 \right\} + O_{\rm even}\left({1\over c^2}\right)+ O\left({1\over
 c^7}\right)\ .
\label{eq:3.20}
\end{equation}

All elements are now in hands in order to compare, in the exterior
near-zone $D_i\cap D_e$, the metrics (\ref{eq:3.1}) and
(\ref{eq:2.17}). The ``source'' multipole moments ${\cal
I}=\{I_L,J_L\}$ are defined by the dominant terms in (\ref{eq:3.19a})
and (\ref{eq:3.19c}),
\begin{mathletters}
\label{eq:3.21}
\begin{eqnarray}
 I_L &\equiv& \int d^3 {\bf x} \left\{ \hat x_L \sigma+ {1\over 2c^2(2l+3)}
  {\bf x}^2 \hat x_L \partial^2_t \sigma  - {4(2l+1)\over c^2(l+1)
  (2l+3)}  \hat x_{iL} \partial_t \sigma_i \right\}
 \ , \label{eq:3.21a} \\
 J_L &\equiv& \int d^3 {\bf x}\,\varepsilon_{ab<i_l} \hat x_{L-1>a} \sigma_b
  \ . \label{eq:3.21b}
\end{eqnarray}
\end{mathletters}
The mass-type moment $I_L$ includes 1PN corrections, while the
current-type moment $J_L$ is Newtonian.  The mass moment $I_L$ was
obtained in \cite{BD89}, where it was shown to parametrize the
asymptotic metric generated by the source at the 1PN order.  When
$l=2$ and $l=3$ we recover the moments introduced in
(\ref{eq:3.7})-(\ref{eq:3.8}).  With (\ref{eq:3.19})-(\ref{eq:3.21}),
the multipole expansions (\ref{eq:3.17}) become
\begin{mathletters}
\label{eq:3.22}
\begin{eqnarray}
 {\cal M} (V^{\rm in}) &=& -c \partial_t \phi^0+ G \sum_{l\geq 0}
  {(-)^l\over l !} \partial_L \left\{ {I_L(t-r/c) +I_L (t+r/c)\over
  2r} \right\} + O_{\rm even}\left({1\over c^4} \right)\ ,
  \label{eq:3.22a}\\ {\cal M}(V^{\rm in}_i) &=&{{c^3}\over 4}
  \partial_i\phi^0 -G \sum_{l\geq 1} {(-)^l\over l !} \partial_{L-1}
  \left\{ {{I^{(1)}}_{iL-1} (t-r/c) +{I^{(1)}}_{iL-1} (t+r/c)\over 2r}
  \right\} \nonumber \\ && -G\sum_{l\geq 1} {(-)^l\over l !} {l\over
  l+1} \varepsilon_{iab}\partial_{aL-1} \left\{ {J_{bL-1} (t-r/c)
  +J_{bL-1} (t+r/c)\over 2r} \right\}\nonumber \\ && + O_{\rm
  even}\left({1\over c^2}\right)+ O\left({1\over c^7}\right)\ .
  \label{eq:3.22b}
\end{eqnarray}
\end{mathletters}
Thus, from the definition (\ref{eq:2.18}) of the external potentials
$V^{\rm ext}_\mu$, we obtain the relationships
\begin{mathletters}
\label{eq:3.23}
\begin{eqnarray}
 {\cal M} (V^{\rm in}) &=& -c \partial_t \phi^0 + V^{\rm ext}[{\cal
 I}] + O_{\rm even}\left({1\over c^4} \right)\ , \label{eq:3.23a}\\
 {\cal M} (V^{\rm in}_i) &=& {{c^3} \over 4} \partial_i \phi^0 +
 V^{\rm ext}_i[{\cal I}] + O_{\rm even}\left({1\over c^2}\right) +
 O\left({1\over c^7}\right) \ , \label{eq:3.23b}
\end{eqnarray}
\end{mathletters}
from which we readily infer that the multipole expansion ${\cal M}
(g^{\rm in}_{\mu\nu})$
of the metric (\ref{eq:3.1}) reads, in $D_i\cap D_e$,
\begin{mathletters}
\label{eq:3.24}
\begin{eqnarray}
 {\cal M}( g^{\rm in}_{00}) +{2 \over c}\partial_t\phi^0 &=& -1
+{2\over c^2} (V^{\rm ext}[{\cal I}] +V^{\rm reac}[{\cal I}]) -
{2\over c^4} (V^{\rm ext}[{\cal I}] +V^{\rm reac}[{\cal I}])^2
\nonumber \\ && \qquad + {1\over c^6} {}_6 {\overline g}_{00}^{\rm in}
+ {1\over c^8} {}_8 {\overline g}_{00}^{\rm in} + O\left({1\over
c^{10}}\right)\ ,\label{eq:3.24a} \\ {\cal M}(g^{\rm in}_{0i}) +
\partial_i \phi^0 &=& -{4\over c^3} (V^{\rm ext}_i[{\cal I}] +V^{\rm
reac}_i[{\cal I}]) + {1\over c^5} {}_5 {\overline g}_{0i}^{\rm in} +
{1\over c^7} {}_7 {\overline g}_{0i}^{\rm in}+ O\left({1\over
c^9}\right)\ ,\label{eq:3.24b} \\ {\cal M}( g^{\rm in}_{ij}) &=&
\delta_{ij} \left[ 1 +{2\over c^2} (V^{\rm ext}[{\cal I}] +V^{\rm
reac}[{\cal I}]) \right] +{1\over c^4} {}_4 {\overline g}_{ij}^{\rm
in} + {1\over c^6} {}_6 {\overline g}_{ij}^{\rm in} +O\left({1\over
c^8}\right) \ .  \label{eq:3.24c}
\end{eqnarray}
\end{mathletters}
Clearly the terms depending on $\phi^0$ have the form of an
infinitesimal gauge transformation of the time coordinate. We can
check that the corresponding coordinate transformation can be treated,
to the considered order, in a linearized way (recall from
(\ref{eq:3.18}) that $\phi^0$ is of order $c^{-3}$).  Finally, in the
``exterior'' coordinates
\begin{mathletters}
\label{eq:3.25}
\begin{eqnarray}
 x_{\rm ext}^0 &=& x^0 +\phi^0 (x^\nu) + O_{\rm even}\left({1\over c^5}\right)
     + O\left({1\over c^9}\right)\ ,\label{eq:3.25a}\\
 x_{\rm ext}^i &=& x^i\ +O_{\rm even}\left({1\over c^4}\right)
  +O\left({1\over c^8}\right)\ ,\label{eq:3.25b}
\end{eqnarray}
\end{mathletters}
the metric (\ref{eq:3.24}) is transformed into the ``exterior'' metric
\begin{mathletters}
\label{eq:3.26}
\begin{eqnarray}
 g^{\rm ext}_{00} &=& -1 +{2\over c^2} (V^{\rm ext}[{\cal I}] +V^{\rm
  reac}[{\cal I}]) - {2\over c^4} (V^{\rm ext}[{\cal I}] +V^{\rm
  reac}[{\cal I}])^2 \nonumber \\ && \qquad + {1\over c^6} {}_6
  \overline g^{\rm ext}_{00} + {1\over c^8} {}_8 \overline g^{\rm
  ext}_{00} + O\left({1\over c^{10}}\right)\ , \label{eq:3.26a} \\
  g^{\rm ext}_{0i} &=& -{4\over c^3} (V^{\rm ext}_i[{\cal I}] +V^{\rm
  reac}_i[{\cal I}]) + {1\over c^5} {}_5 \overline g^{\rm ext}_{0i} +
  {1\over c^7} {}_7 \overline g^{\rm ext}_{0i}+ O\left({1\over
  c^9}\right)\ , \label{eq:3.26b} \\ g^{\rm ext}_{ij} &=& \delta_{ij}
  \left[ 1 +{2\over c^2} (V^{\rm ext}[{\cal I}] +V^{\rm reac}[{\cal
  I}]) \right] + {1\over c^4} {}_4 \overline g^{\rm ext}_{ij} +
  {1\over c^6} {}_6 \overline g^{\rm ext}_{ij}+ O\left({1\over
  c^8}\right) \ .  \label{eq:3.26c}
\end{eqnarray}
\end{mathletters}
This metric is exactly identical, as concerns the 1PN, 2.5PN, and
3.5PN approximations, to the exterior metric (\ref{eq:2.17}) obtained
in paper~I, except that here the metric is parametrized by the known
multipole moments ${\cal I}$ instead of the arbitrary moments
$\widetilde {\cal M}$. Thus, we conclude that the two metrics
(\ref{eq:3.1}) and (\ref{eq:2.17}) match in the overlapping region
$D_i\cap D_e$ if (and only if) there is agreement between both types
of multipole moments.  This determines $\widetilde M_L$ and
$\widetilde S_L$.  More precisely, we find that $\widetilde M_L$ and
$\widetilde S_L$ must be related to $I_L$ and $J_L$ given in
(\ref{eq:3.21}) by
\begin{mathletters}
\label{eq:3.27}
\begin{eqnarray}
 \widetilde M_L &=& I_L + O_{\rm even} \left( {1 \over c^4} \right)
   + O\left({1\over c^8}\right)\ , \label{eq:3.27a} \\
 \widetilde S_L &=& J_L + O_{\rm even} \left( {1 \over c^2} \right)
  + O\left({1\over c^6}\right) \ , \label{eq:3.27b}
\end{eqnarray}
\end{mathletters}
where as usual the relation for $\widetilde M_L$ is accurate to 1PN
order, and the relation for $\widetilde S_L$ is Newtonian (we do also
control the parity of some neglected terms). Satisfying the latter
matching solves the problem at hand, by showing that the inner metric
(\ref{eq:3.1})-(\ref{eq:3.8}) results from the post-Newtonian
expansion of a solution of the (non-linear) field equations and a
condition of no incoming radiation.

We emphasize the dependence of the result on the coordinate system.
Of course, the metric (\ref{eq:3.1}), which contains the reactive
potentials (\ref{eq:3.6})-(\ref{eq:3.8}), is valid only in its own
coordinate system.  It is a well-known consequence of the equivalence
principle that radiation reaction forces in general relativity are
inherently dependent on the coordinate system (see e.g.  \cite{S83}
for a comparison between various expressions of the radiation reaction
force at the Newtonian order).  The coordinate system in which the
reactive potentials (\ref{eq:3.6})-(\ref{eq:3.8}) are valid is defined
as follows.  We start from the particular coordinate system in which
the linearized metric reads (\ref{eq:2.3}).  Then we apply two
successive coordinate transformations.  The first one is associated
with the gauge vector $\xi^\mu$ given by (\ref{eq:2.7}), and the
second one is associated with $\phi^\mu$ whose only needed component
is $\phi^0$ given by (\ref{eq:3.18}).  The resulting coordinate system
is the one in which $V^{\rm reac}_\mu$ is valid.  (Actually, the gauge
vector $\xi^\mu$ should be modified according to the procedure defined
in Sect.  II.C of paper I, so that the good fall-off properties of the
metric at infinity are preserved.)

\section{The balance equations to post-Newtonian order}\label{sec:4}

\subsection{Conservation laws for energy and momenta at 1PN order}

Up to the second post-Newtonian approximation of general relativity,
an isolated system admits some conserved energy, linear momentum, and
angular momentum.  These have been obtained, in the case of weakly
self-gravitating fluid systems, by Chandrasekhar and Nutku
\cite{CN69}.  The less accurate 1PN-conserved quantities were obtained
before, notably by Fock \cite{Fock}.  In this subsection we re-derive,
within the present framework (using in particular the mass density
$\sigma$ defined in (\ref{eq:3.3a})), the 1PN-conserved energy and
momenta of the system.  The 1PN energy and momenta are needed in the
next subsection, in which we establish their laws of variation during
the emission of radiation at 1PN order (hence we do not need the more
occurate 2PN-conserved quantities).

To 1PN order the metric (\ref{eq:3.1}) reduces to
\begin{mathletters}
\label{eq:4.1}
\begin{eqnarray}
 g^{\rm in}_{00} &=& -1 +{2\over c^2} V^{\rm in} - {2\over c^4}
  (V^{\rm in})^2 + O\left({1\over c^{6}}\right)\ , \label{eq:4.1a} \\
  g^{\rm in}_{0i} &=& -{4\over c^3} V^{\rm in}_i + O\left({1\over
  c^5}\right)\ ,\label{eq:4.1b} \\ g^{\rm in}_{ij} &=& \delta_{ij}
  \left(1 +{2\over c^2} V^{\rm in}\right) + O\left({1\over c^4}\right)
  \ , \label{eq:4.1c}
\end{eqnarray}
\end{mathletters}
where $V^{\rm in}$ and $V^{\rm in}_i$ are given by (\ref{eq:3.4}).
In fact, $V^{\rm in}$ and $V^{\rm in}_i$ are given by their
post-Newtonian expansions (\ref{eq:3.5}), which can be limited here
to the terms
\begin{mathletters}
\label{eq:4.2}
\begin{eqnarray}
 V^{\rm in} &=& U + {1\over 2c^2}\, \partial^2_t X
         + O\left({1\over c^4} \right)\ , \label{eq:4.2a} \\
 V^{\rm in}_i &=& U_i + O\left( {1\over c^2} \right)\ , \label{eq:4.2b}
\end{eqnarray}
\end{mathletters}
where the instantaneous (Poisson-like) potentials $U$, $X$ and $U_i$ are 
defined by
\begin{mathletters}
\label{eq:4.3}
\begin{eqnarray}
 U({\bf x},t) &=& G \int {d^3{\bf x}'\over |{\bf x}-{\bf x}'|} \sigma
 ({\bf x}',t)\ , \label{eq:4.3a}\\
 X({\bf x},t) &=& G \int d^3{\bf x}'|{\bf x}-{\bf x}'| \sigma
 ({\bf x}',t)\ , \label{eq:4.3b}\\
 U_i({\bf x},t) &=& G \int {d^3{\bf x}'\over |{\bf x}-{\bf x}'|} \sigma_i
 ({\bf x}',t)\ . \label{eq:4.3c}
\end{eqnarray}
\end{mathletters}
Since $V^{\rm in}$ is a symmetric integral, there are no terms of
order $c^{-3}$ in (\ref{eq:4.2a}) (such a term would be a simple
function of time in the case of a retarded integral).  We shall need
(only in this subsection) a metric whose space-space components $ij$
are more accurate than in (\ref{eq:4.1c}), taking into account the
next-order correction term.  We introduce an instantaneous potential
whose source is the sum of the matter stresses, say $\sigma_{ij} =
T^{ij}$, and the (Newtonian) gravitational stresses,
\begin{mathletters}
\label{eq:4.4}
\begin{equation}
 P_{ij}({\bf x},t) = G \int {d^3{\bf x}'\over |{\bf x}-{\bf x}'|}
  \left[ \sigma_{ij} + {1\over 4\pi G} \left( \partial_i U\partial_j U
 - {1\over 2} \delta_{ij} \partial_k U\partial_k U\right) \right]
  ({\bf x}',t)\ . \label{eq:4.4a}
\end{equation}
The spatial trace $P\equiv P_{ii}$ is
\begin{equation}
P({\bf x},t) = G \int {d^3{\bf x}'\over |{\bf x}-{\bf x}'|}
  \left[ \sigma_{ii} - {1\over 2} \sigma U \right]
  ({\bf x}',t)  +  {{U^2} \over {4}} \ . \label{eq:4.4b}
\end{equation}
\end{mathletters}
The metric which is accurate enough for our purpose reads, in
terms of the instantaneous potentials (\ref{eq:4.3})-(\ref{eq:4.4}),
\begin{mathletters}
\label{eq:4.5}
\begin{eqnarray}
 g^{\rm in}_{00} &=& -1 +{2\over c^2} U + {1\over c^4} [\partial_t^2
   X - 2U^2] + O\left( {1\over c^6} \right)\ , \label{eq:4.5a} \\
 g^{\rm in}_{0i} &=& -{4\over c^3} U_i
 + O\left( {1\over c^5} \right) \label{eq:4.5b}\ , \\
 g^{\rm in}_{ij} &=& \delta_{ij}
 \left( 1 +{2\over c^2} U + {1\over c^4} [\partial^2_t X +2U^2] \right)
 + {4\over c^4} [P_{ij} -\delta_{ij} P] +
 O\left( {1\over c^6} \right)\ . \label{eq:4.5c}
\end{eqnarray}
The square-root of (minus) the determinant of the metric is
\begin{equation}
\sqrt{-g^{\rm in}} = 1 + {2\over c^2} U + {1\over c^4} [\partial^2_t X
+2U^2 - 4P] + O \left( {1\over c^6} \right)\ .  \label{eq:4.5d}
\end{equation}
\end{mathletters}
Consider the local equations of motion of the source, which state the
conservation in the covariant sense of the stress-energy tensor $T^{\mu\nu}$ 
(i.e.  $\nabla_\mu T^\mu_\alpha=0$). These equations, written in a form
adequate for our purpose, are
\begin{equation}
 \partial_\mu \Pi^\mu_\alpha = {\cal F}_\alpha \ , \label{eq:4.6}
\end{equation}
where the left-hand-side is the divergence in the ordinary sense of the
material stress-energy density
\begin{equation}
 \Pi_\alpha^\mu \equiv \sqrt{-g^{\rm in}}\ g^{\rm in}_{\alpha\nu}
    T^{\mu\nu} \ , \label{eq:4.7}
\end{equation}
and where the right-hand-side can be viewed as the four-force density   
\begin{equation}
 {\cal F}_\alpha \equiv {1\over 2}
 \sqrt{-g^{\rm in }}\ T^{\mu\nu} \partial_\alpha
  g^{\rm in}_{\mu\nu}\  \label{eq:4.8}
\end{equation}
The 1PN-conserved energy and momenta follow from integration of these
equations over the ordinary three-dimensional space, which yields the
following three laws (using the Gauss theorem to discard some
divergences of compact-supported terms)
\begin{mathletters}
\label{eq:4.9}
\begin{eqnarray}
 {d \over dt}\left\{- \int d^3{\bf x}\, \Pi^0_0\right\} &=& -c \int 
  d^3{\bf x}\,{\cal F}_0\ , \label{eq:4.9a}\\
 {d \over dt}\left\{{1\over c}\int d^3{\bf x}\,\Pi^0_i\right\} &=& \int 
  d^3{\bf x}\,{\cal F}_i\ , \label{eq:4.9b}\\
 {d \over dt}\left\{{1\over c} \varepsilon_{ijk} \int d^3{\bf x}\, x_j
  \Pi^0_k\right\} &=& \varepsilon_{ijk}
  \int d^3{\bf x}\,\bigl( x_j {\cal F}_k +\Pi_k^j \bigr) \ .\label{eq:4.9c}
\end{eqnarray}
\end{mathletters}
The quantities $\Pi_\alpha^\mu$ and ${\cal F}_\alpha$ are then determined.
With (\ref{eq:4.5}) we
obtain, for the various components of $\Pi_\alpha^\mu$,
\begin{mathletters}
\label{eq:4.10}
\begin{eqnarray}
 \Pi^0_0 &=& -\sigma c^2 + \sigma_{ii} + {4\over c^2} [\sigma
   P-\sigma_i U_i] + O\left( {1\over c^4} \right)\ ,
   \label{eq:4.10a}\\ \Pi^0_i &=& c \sigma_i \left( 1+4 {U\over c^2}
   \right) - {4\over c} \sigma U_i + O\left( {1\over c^3} \right)\ ,
   \label{eq:4.10b}\\ \Pi^i_0 &=& -c \sigma_i \left( 1-4 {P\over c^4}
   \right) - {4\over c^3} \sigma_{ij} U_j + O\left( {1\over c^5}
   \right)\ , \label{eq:4.10c}\\ \Pi^i_j &=& \sigma_{ij} \left( 1+4
   {U\over c^2} \right) - {4\over c^2} \sigma_i U_j + O\left( {1\over
   c^4} \right)\ . \label{eq:4.10d}
\end{eqnarray}
\end{mathletters}
Note that $\Pi_0^i$ is determined with a better precision than
$\Pi_i^0$ (but we shall not need this higher precision and give it for
completeness).  For the components of ${\cal F}_\alpha$, we find
\begin{mathletters}
\label{eq:4.11}
\begin{eqnarray}
 {\cal F}_0 &=& {1\over c} \sigma \partial_t \left( U +{1\over 2c^2}
  \partial_t^2 X\right) - {4\over c^3} \sigma_j \partial_t U_j +
  O \left( {1\over c^5} \right)\ , \label{eq:4.11a} \\
 {\cal F}_i &=& \sigma \partial_i \left( U +{1\over 2c^2}
  \partial_t^2 X\right) - {4\over c^2} \sigma_j \partial_i U_j +
  O \left( {1\over c^4} \right)\ . \label{eq:4.11b}
\end{eqnarray}
\end{mathletters}
The $\Pi_\alpha^\mu$'s and ${\cal F}_\alpha$'s can now be substituted
into the integrals on both sides of (\ref{eq:4.9}). This is correct
because the support of the integrals is the compact support of the
source, which is, for a slowly-moving source, entirely located within
the source's near-zone $D_i$, where the post-Newtonian expansion is
valid.  Straightforward computations permit us to re-express the
right-hand-sides of (\ref{eq:4.9}) into the form of total
time-derivatives.  We do not detail here this computation which is
well-known (at 1PN order), but we present in Sect.~IV.B a somewhat
general formula which can be used to reach elegantly the result (see
Eq.~(4.23)).  By transfering the total time-derivatives to the
left-hand-sides of (\ref{eq:4.9}), one obtains the looked-for
conservation laws at 1PN order, namely
\begin{eqnarray}
 {d E^{\rm 1PN}\over dt} &=& O\left({1\over c^{4}}
   \right)\ ,\label{eq:4.14} \\
 {d P_i^{\rm 1PN}\over dt} &=& O\left(
   {1\over c^{4}}\right)\ , \label{eq:4.15} \\
 {d S_i^{\rm 1PN}\over dt} &=& O\left(
   {1\over c^{4}}\right)\ , \label{eq:4.16}
\end{eqnarray}
where the 1PN energy $E^{\rm 1PN}$, linear momentum $P_i^{\rm 1PN}$,
and angular momentum $S_i^{\rm 1PN}$ are given by the
integrals over the source
\FL
\begin{eqnarray}
  E^{\rm 1PN} &=& \int d^3 {\bf x} \left\{ \sigma c^2 + {1\over 2}
  \sigma U - \sigma_{ii} + {1\over c^2} \biggl[ -4\sigma P + 2\sigma_i
  U_i + {1\over 2} \sigma \partial^2_t X -{1\over 4} \partial_t \sigma
  \partial_t X \biggr] \right\}\ , \label{eq:4.17} \\ P^{\rm 1PN}_i
  &=& \int d^3 {\bf x} \left\{ \sigma_i - {1\over 2c^2} \sigma
  \partial_i \partial_t X \right\}\ , \label{eq:4.18} \\ S^{\rm 1PN}_i
  &=& \varepsilon_{ijk} \int d^3 {\bf x} x_j \left\{ \sigma_k +
  {1\over c^2} \left[ 4 \sigma_k U - 4\sigma U_k - {1\over 2} \sigma
  \partial_k \partial_t X \right] \right\}\ . \label{eq:4.19}
\end{eqnarray}
The 1PN energy $E^{\rm 1PN}$ can also be written as \cite{R}
\FL
\begin{eqnarray}
  E^{\rm 1PN}&=& \int d^3 {\bf x} \left\{ \sigma c^2 + {1\over 2} \sigma U
   - \sigma_{ii} + {1\over c^2} \left[ \sigma U^2 -4\sigma_{ii} U
   +2 \sigma_i U_i
   + {1\over 2} \sigma \partial^2_t X
    - {1\over 4} \partial_t \sigma \partial_t X \right] \right\}\ .
  \nonumber \\
  \label{eq:4.15'}
\end{eqnarray}
A similar but more precise computation would yield the 2PN-conserved
quantities \cite{CN69}.

\subsection{Secular losses of the 1PN-accurate energy and momenta}

As the reactive potentials $V^{\rm reac}_\mu$ manifestly change sign
in a time reversal, they are expected to yield dissipative effects in
the dynamics of the system, i.e.  secular losses of its total energy,
angular momentum and linear momentum.  The ``Newtonian'' radiation
reaction force is known to extract energy in the system at the same
rate as given by the Einstein quadrupole formula, both in the case of
weakly self-gravitating systems
\cite{Bu69,Th69,Bu71,C69,CN69,CE70,EhlRGH,WalW80,BD84,AD75,PaL81,Ehl80,Ker80,Ker80',BRu81,BRu82,S85}
and compact binary systems \cite{DD81a,DD81b,D82}.  Similarly the
reaction force extracts angular momentum in the system.  As concerns
linear momentum the Newtonian reaction force is not precise enough, and
one needs to go to 1PN order.

In this subsection, we prove that the 1PN-accurate reactive potentials
$V^{\rm reac}_\mu$ lead to decreases of the 1PN-accurate energy and
momenta (computed in (\ref{eq:4.17})-(\ref{eq:4.15'})) which are in
perfect agreement with the corresponding far-zone fluxes, known from
the works \cite{EW75,Th80,BD89} in the case of the energy and angular
momentum, and from the works
\cite{Pa71,Bek73,Press77,Th80} in the case of linear momentum.

We start again from the equations of motion
(\ref{eq:4.6})-(\ref{eq:4.8}), which imply, after spatial integration,
the laws (\ref{eq:4.9}) that we recopy here:
\begin{mathletters}
\label{eq:4.20}
\begin{eqnarray}
 {d \over dt}\left\{- \int d^3{\bf x}\, \Pi^0_0\right\} &=& -c \int 
  d^3{\bf x}\,{\cal F}_0\ , \label{eq:4.20a}\\
 {d \over dt}\left\{{1\over c}\int d^3{\bf x}\,\Pi^0_i\right\} &=& \int 
  d^3{\bf x}\,{\cal F}_i\ , \label{eq:4.20b}\\
 {d \over dt}\left\{{1\over c} \varepsilon_{ijk} \int d^3{\bf x}\, x_j
  \Pi^0_k\right\} &=& \varepsilon_{ijk}
   \int d^3{\bf x}\,\bigl( x_j {\cal F}_k +\Pi_k^j \bigr) \ .\label{eq:4.20c}
\end{eqnarray}
\end{mathletters}
The left-hand-sides are in the form of total time-derivatives. To 1PN
order, we have seen that the right-hand-sides can be transformed into
total time-derivatives, which combine with the left-hand-sides to give
the 1PN-conserved energy and momenta.  Here we shall prove that the
contributions due to the reactive potentials in the right-hand-sides
cannot be transformed entirely into total time-derivatives, and that
the remaining terms yield precisely the corresponding 1PN fluxes. The
balance equations then follow (modulo a slight assumption and a
general argument stated below).

The right-hand-sides of (\ref{eq:4.20}) are evaluated by substituting
the metric (\ref{eq:3.1}), involving the potentials ${\cal V}^{\rm in}_\mu
= V^{\rm in}_\mu+ V^{\rm reac}_\mu$.  The components of the force
density (\ref{eq:4.8}) are found to be
\begin{mathletters}
\label{eq:4.21}
\begin{eqnarray}
 {\cal F}_0 &=& {\sigma\over c} \partial_t {\cal V}^{\rm in}
   - {4\over c^3} \sigma_j
  \partial_t {\cal V}^{\rm in}_j + 
{1 \over c^{5}} {}_5 {\cal F}_0 + {1 \over c^{7}}
{}_7 {\cal F}_0 + O\left({1\over c^{9}}\right) \ ,
  \label{eq:4.21a} \\
 {\cal F}_i &=& \sigma \partial_i {\cal V}^{\rm in} - {4\over c^2} \sigma_j
   \partial_i {\cal V}^{\rm in}_j + {1 \over c^{4}} {}_4 {\cal F}_i
   + {1 \over c^{6}}
{}_6 {\cal F}_i + O\left({1\over c^{8}}\right) \ ,  \label{eq:4.21b}
\end{eqnarray}
\end{mathletters}
where we have been careful at handling correctly the un-controlled 2PN
and 3PN approximations, which lead to the terms symbolized by the
$c^{-n}{}_n {\cal F}_{\mu}$'s.  The equations (\ref{eq:4.21}) reduce
to (\ref{eq:4.11}) at the 1PN approximation.  They give the components
of the force as linear functionals of ${\cal V}^{\rm in}$ and ${\cal
V}^{\rm in}_i$.  The remainders are 4PN at least.
The same
computation using the stress-energy density (\ref {eq:4.7}) yields the
term which is further needed in (\ref{eq:4.20c}),
\begin{equation}
  \varepsilon_{ijk} \Pi^j_k = - {4\over c^2} \varepsilon_{ijk} \sigma_j
 {\cal V}^{\rm in}_k +{1 \over c^4} {}_4 {\cal T}_i
  + {1 \over c^{6}} {}_6 {\cal T}_i
  + O\left({1\over c^{8}}\right) \ . \label{eq:4.22}
\end{equation}
The ${}_n {\cal T}_i$'s represent the 2PN and 3PN
approximations. Thanks to (\ref{eq:4.21})-(\ref{eq:4.22}) one can now
transform the laws (\ref{eq:4.20}) into
\begin{mathletters}
\label{eq:4.23}
\begin{eqnarray}
 {d \over dt}\left\{- \int d^3{\bf x}\, \Pi^0_0\right\} 
   &=& \int d^3{\bf x}\,\left\{ -\sigma \partial_t {\cal V}^{\rm in}
   + {4\over c^2} \sigma_j \partial_t {\cal V}^{\rm in}_j \right\}
   + {1 \over c^{4}} {}_4 X + {1 \over c^{6}} {}_6 X + O\left({1\over c^{8}}
   \right)\ ,\nonumber \\ \label{eq:4.23a} \\
 {d \over dt}\left\{{1\over c}\int d^3{\bf x}\,\Pi^0_i\right\} 
   &=& \int d^3{\bf x}\,\left\{ \sigma \partial_i
   {\cal V}^{\rm in} - {4\over c^2} \sigma_j \partial_i {\cal V}^{\rm in}_j 
\right\}
   + {1 \over c^{4}} {}_4 Y_i + {1 \over c^{6}} {}_6 Y_i + O\left(
   {1\over c^{8}}\right)\ , \nonumber \\ \label{eq:4.23b} \\
 {d \over dt}\left\{{1\over c} \varepsilon_{ijk} \int d^3{\bf x}\, x_j
  \Pi^0_k\right\} &=& \varepsilon_{ijk} \int d^3{\bf x}\,
   \left\{ \sigma x_j \partial_k {\cal V}^{\rm in} 
   - {4\over c^2} \sigma_m x_j \partial_k {\cal V}^{\rm in}_m - {4\over c^2}
   \sigma_j {\cal V}^{\rm in}_k \right\} \nonumber\\
 &&\qquad\qquad+ {1 \over c^{4}} {}_4 Z_i +{1 \over c^{6}}{}_6Z_i + O\left(
{1\over c^{8}}\right)\ , \label{eq:4.23c}
\end{eqnarray}
\end{mathletters}
where ${}_n X$, ${}_n Y_i$, and ${}_n Z_i$ denote some spatial
integrals of the 2PN and 3PN terms in (\ref{eq:4.21})-(\ref{eq:4.22}).

Consider first the piece in ${\cal V}^{\rm in}_\mu$ which is composed of
the potential $V_\mu^{\rm in}$, given by the symmetric integral
(\ref{eq:3.4}) or by the Taylor expansion (\ref{eq:3.5}).  To 1PN order
$V^{\rm in}_\mu$ contributes to the laws (\ref{eq:4.23})
only in the form of total time-derivatives (see Sect IV.A).  Here we
present a more general proof of this result, valid formally up to any
post-Newtonian order. This proof shows that the result is due to the very
structure of the symmetric potential $V_\mu^{\rm in}$ as given by
(\ref{eq:3.5}).  The technical formula sustaining the
proof is
\begin{equation}
 \sigma_\mu (x) \partial^N_t \sigma_\nu (x') + (-)^{N+1} \sigma_\nu
 (x') \partial^N_t \sigma_\mu (x) = {d\over dt} \left\{ \sum^{N-1}_{q=0}
 (-)^q \partial^q_t \sigma_\mu (x) \partial^{N-q-1}_t \sigma_\nu (x')
   \right\}\ , \label{eq:4.24}
\end{equation}
where $x \equiv ({\bf x},t)$ and $x'\equiv ({\bf x}',t)$ denote two
field points (located in the same hypersurface $t \equiv x^0/c
=$~const), and where $N$ is some integer [we recall the notation
$\sigma_\mu \equiv (\sigma,\sigma_i)$]. The contributions of the
symmetric potential in the energy and linear momentum laws
(\ref{eq:4.23a}) and (\ref{eq:4.23b}) are all of the same type,
involving the spatial integral of $\sigma_\mu (x) \partial_{\alpha}
V^{\rm in}_\mu (x)$ (with no summation on $\mu$ and $\alpha =0,i$).
One replaces into this spatial integral the potential $V^{\rm in}_\mu$
by its Taylor expansion when $c \to \infty$ as given by
(\ref{eq:3.5}).  This yields a series of terms involving when the
index $\alpha = 0$ the double spatial integral of $|{\bf x}-{\bf
x}'|^{2p-1} \sigma_\mu (x) \partial^{2p+1}_t \sigma_\mu (x')$, and
when $\alpha = i$ the double integral of $\partial_i |{\bf x}-{\bf
x}'|^{2p-1}
\sigma_\mu (x) \partial^{2p}_t \sigma_\mu (x')$. By symmetrizing the
integrand under the exchange ${\bf x} \leftrightarrow {\bf x}'$ and by
using the formula (\ref{eq:4.24}) where $N = 2p+1$ when $\alpha = 0$ and
$N = 2p$ when $\alpha = 1,2,3$, one finds that the integral is indeed a
total time-derivative. The same is true for the symmetric contributions in
the angular momentum law (\ref{eq:4.23c}), which yield a series of integrals of
$\varepsilon
_{ijk} x_j \partial_k |{\bf x}-{\bf x}'|^{2p-1} \sigma_\mu (x)
\partial^{2p}_t \sigma_\mu (x')$ and $\varepsilon_{ijk} |{\bf x}-{\bf
x}'|^{2p-1} \sigma_j (x) \partial^{2p}_t \sigma_k (x')$, on which one
uses the formula (\ref{eq:4.24}) where $N = 2p$.  

Thus the symmetric (inner) potentials $V_\mu^{\rm in}$ contribute to
the right-hand-sides of (\ref{eq:4.23}) only in the form of total
time-derivatives. This is true even though, as noticed earlier, the
potentials $V^{\rm in}_\mu$ contain some reactive (``time-odd'')
terms, through the contributions of the source densities
$\sigma_\mu$. As shown here, these time-odd terms combine with the
other time-odd terms present in the $\sigma_\mu$'s appearing
explicitly in (\ref{eq:4.23}) to form time-derivatives.  Such time-odd
terms will not participate ultimately to the balance equations, but
they do participate to the complete 3.5PN approximation in the
equations of motion of the system.  This fact has been noticed, and
these time-odd terms computed for binary systems, by Iyer and Will
\cite{IW95} (see their equations (\ref{eq:3.8}) and (\ref{eq:3.9})).

The numerous time-derivatives resulting from the symmetric potentials
are then transferred to the left-hand-sides of (\ref{eq:4.23}). To 1PN
order these time-derivatives permit reconstructing the 1PN-conserved
energy and momenta $E^{\rm 1PN}$, $P_i^{\rm 1PN}$ and $S_i^{\rm
1PN}$. We include also the time-derivatives of higher order but they
will be negligible in the balance equations (see below).  Therefore,
we have proved that the laws (\ref{eq:4.23}) can be re-written as
\begin{mathletters}
\label{eq:4.25}
\begin{eqnarray}
{d\over dt} \left[ E^{\rm 1PN}+{\overline O}\left({1\over c^{4}}
    \right)\right] &=& \int d^3{\bf x}\,\left\{ -\sigma \partial_t
 {V}^{\rm reac}+ {4\over c^2} \sigma_j \partial_t {V}^{\rm reac}_j \right\}
 + {1 \over c^{4}} {}_4 X + {1 \over c^{6}} {}_6 X + O\left({1\over c^{8}}
    \right)\ , \nonumber \\        \label{eq:4.25a} \\
{d\over dt} \left[ P^{\rm 1PN}_i +{\overline O}\left({1\over c^{4}}
    \right)\right] &=& \int d^3{\bf x}\,\left\{ \sigma
 \partial_i {V}^{\rm reac} -{4\over c^2} \sigma_j \partial_i {V}^{\rm reac}_j
  \right\} + {1 \over c^{4}} {}_4 Y_i + {1 \over c^{6}} {}_6 Y_i +
  O\left({1\over c^{8}}\right)\ , \nonumber \\      \label{eq:4.25b} \\
 {d\over dt} \left[ S^{\rm 1PN}_i +{\overline O}\left({1\over c^{4}}
    \right)\right] &=& \varepsilon_{ijk} \int d^3{\bf x}\,
   \left\{ \sigma x_j \partial_k {V}^{\rm reac}
   - {4\over c^2} \sigma_m x_j \partial_k {V}^{\rm reac}_m - {4\over c^2}
    \sigma_j {V}^{\rm reac}_k \right\} \nonumber\\
 &&\qquad\qquad+ {1 \over c^{4}} {}_4 Z_i + {1 \over c^{6}} {}_6 Z_i + O
    \left({1\over c^{8}}\right)\ .       \label{eq:4.25c}
\end{eqnarray}
\end{mathletters}
The $O$-symbols $\overline O (c^{-4})$ denote the terms, coming in
particular from the symmetric potentials in the right-hand-sides,
which are of higher order than 1PN.  We add an overbar on these
remainder terms to distinguish them from other terms introduced below.

Now recall
that the 2PN and 3PN approximations, including in particular the terms
${}_n X$, ${}_n Y_i$, and ${}_n Z_i$ in (\ref{eq:4.25}),
are non-radiative (non-dissipative).  Indeed they correspond to ``even''
approximations, and depend instantaneously on the parameters of the
source.  In the case of the 2PN approximation, Chandrasekhar and Nutku
\cite{CN69} have proved explicitly that ${}_4X$, ${}_4Y_i$,
and ${}_4Z_i$ can be transformed into total time-derivatives, leading
to the expressions of the 2PN-conserved energy and momenta.  Here we
shall assume that the same property holds for the 3PN approximation,
namely that the terms ${}_6X$, ${}_6Y_i$, and ${}_6Z_i$ can also be
transformed into time-derivatives.  This assumption is almost
certainly correct.  The 3PN approximation is not expected to yield any
secular decrease of quasi-conserved quantities.  It can be argued, in
fact, that the 3PN approximation is the last approximation which is
purely non-dissipative.  Under this (slight) assumption we can now
transfer the terms ${}_n X$, ${}_n Y_i$, and ${}_n Z_i$ to the
left-hand-sides, where they modify the remainder terms $\overline O
(c^{-4})$.  Thus,
\begin{mathletters}
\label{eq:4.26}
\begin{eqnarray}
{d\over dt} \left[ E^{\rm 1PN}+{\widetilde O}\left({1\over c^{4}}
    \right)\right] &=& \int d^3 {\bf x} \left\{ -\sigma \partial_t
  V^{\rm reac} + {4\over c^2} \sigma_j \partial_t V^{\rm reac}_j \right\}
  + O \left( {1\over c^8}\right)\ , \label{eq:4.26a}\\
{d\over dt} \left[ P^{\rm 1PN}_i +{\widetilde O}\left({1\over c^{4}}
    \right)\right] &=& \int d^3 {\bf x}\left\{ \sigma \partial_i
  V^{\rm reac} - {4\over c^2} \sigma_j \partial_i V^{\rm reac}_j \right\}
  + O \left( {1\over c^8}\right)\ , \label{eq:4.26b}\\
{d\over dt} \left[ S^{\rm 1PN}_i +{\widetilde O}\left({1\over c^{4}}
    \right)\right] &=& \varepsilon_{ijk}
  \int d^3 {\bf x} \left\{ \sigma x_j \partial_k V^{\rm reac}
  - {4\over c^2} \sigma_m x_j \partial_k V^{\rm reac}_m
  - {4\over c^2} \sigma_j V^{\rm reac}_k \right\} \nonumber  \\
  && + O \left( {1\over c^8}\right)\ , \label{eq:4.26c}
\end{eqnarray}
\end{mathletters}
where $\widetilde O(c^{-4})$ denotes the modified remainder
terms, which satisfy, for instance,
$E^{1PN}+\widetilde O(c^{-4}) = E^{2PN}+ O(c^{-5})$.

The equations (\ref{eq:4.26}) clarify the way the losses of energy and
momenta are driven by the radiation reaction potentials.  However,
these equations are still to be transformed using the explicit
expressions (\ref{eq:3.6})-(\ref{eq:3.8}).  When inserting these
expressions into the right-hand-sides of (\ref{eq:4.26}) one is left
with numerous terms.  All these terms have to be transformed and
combined together modulo total time-derivatives.  Thus, numerous
operations by parts on the time variable are performed
[i.e. $A\partial_t B =\partial_t (AB) -B\partial_t A$], thereby
producing many time-derivatives which are transferred as before to the
left-hand-sides of the equations, where they modify the ${\widetilde
O} (c^{-4})$'s by some contributions of order $c^{-5}$ at least (since
this is the order of the reactive terms).  During the transformation
of the laws (\ref{eq:4.26a}) and (\ref{eq:4.26c}) for the energy and
angular momentum, it is crucial to recognize among the terms the
expression of the 1PN-accurate mass quadrupole moment $I_{ij}$ given
by (\ref{eq:3.7}) (or (\ref{eq:3.21a}) with $l = 2$), namely
\begin{equation}
 I_{ij} = \int d^3{\bf x} \left\{ \hat x_{ij} \sigma + {1\over 14c^2}
 {\bf x}^2 \hat x_{ij} \partial_t^2 \sigma  - {20\over 21c^2} \hat x_{ijk}
\partial_t \sigma_k  \right\}\ .\label{eq:4.27}
\end{equation}
And during the transformation of the law (\ref{eq:4.26b}) for linear
momentum, the important point is to remember that the 1PN-accurate
mass dipole moment $I_i$, whose second time-derivative is zero as a
consequence of the equations of motion $[d^2I_i/dt^2 =O(c^{-4})]$,
reads
\begin{mathletters}
\label{eq:4.28}
\begin{equation}
 I_i = \int d^3 {\bf x} \left\{ x_i \sigma + {1\over 10c^2}{\bf x}^2 x_i
\partial^2_t \sigma - {6\over 5c^2}  \hat x_{ij} \partial_t \sigma_j
   \right\}\ . \label{eq:4.28a}
\end{equation}
This moment is also a particular case, when $l = 1$, of the general
formula (\ref{eq:3.21a}). An alternative expression of the dipole moment is
\begin{equation}
 I_i = \int d^3 {\bf x}~ x_i \left\{ \sigma + {1\over c^2} \left(
 {1\over 2}\sigma U - \sigma_{jj} \right) \right\} + O\left( {1\over
 c^4}\right) \ ,  \label{eq:4.28b}
\end{equation}
\end{mathletters}
which involves the Newtonian conserved mass density (given also by the
three first terms in (4.15)). Finally, the end results of the
computations are some laws involving in the right-hand-sides some
quadratic products of derivatives of multipole moments, most
importantly the 1PN quadrupole moment (\ref{eq:4.27}) (the other
moments being Newtonian), namely
\begin{mathletters}
\label{eq:4.29}
\begin{eqnarray}
{d\over dt} \left[ E^{\rm 1PN}+{\widehat O}\left({1\over c^{4}}
    \right)\right] &=& - {G\over c^5} \left\{ {1\over 5} I^{(3)}_{ij}
  I^{(3)}_{ij} + {1\over c^2}\left[{1\over 189} I^{(4)}_{ijk} I^{(4)}_{ijk}
  + {16\over 45} J^{(3)}_{ij} J^{(3)}_{ij} \right] \right\}
  + O\left( {1\over c^8}\right) \ , \nonumber \\   \label{eq:4.29a} \\
{d\over dt} \left[ P^{\rm 1PN}_i +{\widehat O}\left({1\over c^{4}}
    \right)\right] &=& - {G\over c^7} \left\{ {2\over 63} I^{(4)}_{ijk}
  I^{(3)}_{jk} + {16\over 45} \varepsilon_{ijk} I^{(3)}_{jm} J^{(3)}_{km}
  \right\}   + O\left( {1\over c^9}\right)\ , \label{eq:4.29b} \\
  {d\over dt} \left[ S^{\rm 1PN}_i +{\widehat O}\left({1\over c^{4}}
    \right)\right] &=& - {G\over c^5} \varepsilon_{ijk} \left\{ {2\over 5}
  I^{(2)}_{jm} I^{(3)}_{km}+{1\over c^2}\left[{1\over 63} I^{(3)}_{jmn}
  I^{(4)}_{kmn} + {32\over 45} J^{(2)}_{jm} J^{(3)}_{km} \right] \right\}
  + O\left( {1\over c^8}\right)\ . \nonumber \\  \label{eq:4.29c}
\end{eqnarray}
\end{mathletters}
The remainders in the left-hand-sides are such that $\widehat
O(c^{-4}) = \widetilde O(c^{-4})+O(c^{-5})$.  The remainders in the
right-hand-sides are $O(c^{-8})$ in the cases of energy and angular
momentum because of tail contributions (see Sect.~IV. C), but is
$O(c^{-9})$ in the case of the linear momentum.

The last step is to argue that the unknown terms in the
left-hand-sides, namely the total time-derivatives of the remainders
${\widehat O}( c^{-4})$, are negligible as compared to the controlled terms
in the right-hand-sides, despite their larger formal post-Newtonian order
($c^{-4}$ vs $c^{-5}$ and $c^{-7}$).  When computing, for instance, the
time evolution of the orbital phase of inspiralling compact binaries
\cite{3mn,FCh93,CF94,P93,CFPS93,TNaka94,Sasa94,TSasa94,P95,BDI95,BDIWW95,WWi96,B96pn},
one uses in the left-hand-side of the balance equation the energy
valid at the {\it same} post-Newtonian order as the energy flux in the
right-hand-side.  Because the difference between the orders of
magnitude of the two sides of the equations is $c^{-5}$, we need to
show that the time-derivative increases the formal post-Newtonian
order by a factor $c^{-5}$.  In (\ref{eq:4.28}) this means $d
{\widehat O}(c^{-4})/dt = O(c^{-9})$ [actually, $O(c^{-8})$ would be
sufficient in (4.28), but $O(c^{-9})$ will be necessary in
Sect.~IV.C].  In the case of inspiralling compact binaries, such an
equation is clearly true, because the terms ${\widehat O}(c^{-4})$
depend only on the orbital separation between the two bodies (the
orbit being circular), and thus depend only on the energy which is
conserved at 2PN order (for non-circular orbits one would have also a
dependence on the angular momentum).  Thus the time-derivative adds,
by the law of composition of derivatives, an extra factor $c^{-5}$
coming from the time-derivative of the energy itself.  More generally,
this would be true for any system whose 2PN dynamics can be
parametrized by the 2PN-conserved energy and angular momentum.  This
argument could perhaps be extended to systems whose 2PN dynamics is
integrable, in the sense that the solutions are parametrized by some
finite set of integrals of motion, including the integral of energy.
Another argument, which is often presented (see e.g.  \cite{BRu81}),
is that the terms $d {\widehat O}\left({c^{-4}}\right)/dt$ are
negligible when taken in average for quasi-periodic systems, for
instance a binary system moving on a quasi-Keplerian orbit.  The time
average of a total time-derivative is clearly numerical small for such
systems, but it seems difficult to quantify precisely the gain in
order of magnitude which is achieved in this way, for general systems.
The most general argument, valid for any system, is that the terms
$d\widehat O(c^{-4})/dt$ are numerically small when one looks at the
evolution of the system over long time scales, for instance $\Delta
t\gg \widehat O(c^{-4}) (dE^{\rm 1PN}/dt)^{-1}$ (see Thorne
\cite{Th83}, p. 46).

Adopting here $d\widehat O(c^{-4})/dt = O(c^{-9})$ and the latter
general argument, we can neglect the terms $\widehat O(c^{-4})$ and
arrive to the 1PN energy-momenta balance equations
\begin{eqnarray}
{{d E^{\rm 1PN}} \over {dt}} &=& - {G\over c^5} \left\{ {1\over 5}
  I^{(3)}_{ij} I^{(3)}_{ij} + {1\over c^2}\left[{1\over 189}
  I^{(4)}_{ijk} I^{(4)}_{ijk} + {16\over 45} J^{(3)}_{ij} J^{(3)}_{ij}
  \right] \right\} + O\left( {1\over c^8}\right) \ , \label{eq:4.29*}
  \\ {{d P^{\rm 1PN}_i} \over {dt}} &=& - {G\over c^7} \left\{ {2\over
  63} I^{(4)}_{ijk} I^{(3)}_{jk} + {16\over 45} \varepsilon_{ijk}
  I^{(3)}_{jm} J^{(3)}_{km} \right\} + O\left( {1\over c^9}\right)\ ,
  \label{eq:4.30} \\ {{d S^{\rm 1PN}_i} \over {dt}} &=& - {G\over c^5}
  \varepsilon_{ijk} \left\{ {2\over 5} I^{(2)}_{jm}
  I^{(3)}_{km}+{1\over c^2}\left[{1\over 63} I^{(3)}_{jmn}
  I^{(4)}_{kmn} + {32\over 45} J^{(2)}_{jm} J^{(3)}_{km} \right]
  \right\} + O\left( {1\over c^8}\right)\ , \label{eq:4.31}
\end{eqnarray}
relating the 1PN-conserved energy and momenta, given by the explicit
integrals over the source (\ref{eq:3.17})-(\ref{eq:3.19}), to some
combinations of derivatives of multipole moments, also given by
explicit integrals over the source (see
(\ref{eq:3.7})-(\ref{eq:3.8})). Note that at this order both sides of
the equations are in the form of compact-support integrals.  The
right-hand-sides of (\ref{eq:4.29*})-(\ref{eq:4.31}) agree exactly
with (minus) the fluxes of energy and momenta as computed in the wave
zone of the system.  See for instance the equations (4.16'), (4.20')
and (4.23') in \cite{Th80}, when truncated to 1PN order [and recalling
that the quadrupole moment which enters the 1PN fluxes is precisely
the one given by (4.26)].  Thus, we can conclude on the validity of
the balance equations at 1PN order, for weakly self-gravitating
systems.

These equations could also be recovered, in principle, from the
relations (\ref{eq:2.14})-(\ref{eq:2.15}) (which were obtained in
paper I).  Indeed (2.14)-(2.15) involve, besides some instantaneous
contributions such as $T_L(t)$, some non-local (or hereditary)
contributions contained in the functions $m(t)$, $m_i(t)$ and
$s_i(t)$.  These contributions modify the constant monopole and dipole
moments $M$, $M_i$ and $S_i$ by some expressions which correspond
exactly to the emitted fluxes.  The balance equations could be
recovered (with, though, less precision than obtained in this paper)
by using the constancy of the monopole and dipoles $M$, $M_i$ and
$S_i$ in the equations (2.14) written for $l=0$ and $l=1$, and by
using the matching equations obtained in (\ref{eq:3.27}), also written
for $l=0$ and $l =1$. Related to this, notice the term involving a
single time-antiderivative in the function $m_i(t)$ of
(\ref{eq:2.15b}), and which is associated with a secular displacement
of the center of mass position.

\subsection{Tail effects at 1.5PN order}

To 1.5PN order in the radiation reaction force appears an hereditary
integral (i.e. an integral extending on the whole past history of the
source), which is associated physically with the effects of
gravitational-wave tails. More precisely, it is shown in \cite{BD88},
using the same combination of approximation methods as used in paper I
and this paper, that the dominant hereditary contribution in the inner
post-Newtonian metric $g^{\rm in}_{\mu \nu}$ (valid all over $D_i$)
arises at the 4PN order.  At this order, the dynamics of a
self-gravitating system is thus intrinsically dependent on the full
past evolution of the system.

In a particular gauge (defined in \cite{BD88}), the 4PN-hereditary contribution
in $g^{\rm in}_{\mu \nu}$ is entirely located in the 00 component of the
metric, and reads
\begin{equation}
g^{\rm in}_{00}|_{\rm hereditary} = -{8G^2M \over 5c^{10}} x^ix^j
\int^{+\infty}_0 d\lambda~ {\rm \ln} \left({\lambda \over 2} \right)
I^{(7)}_{ij} (t-\lambda) + O\left({1\over c^{11}}\right)\ . \label{eq:4.32}
\end{equation}
The hereditary contributions in the other components of the metric
($0i$ and $ij$) arise at higher order.  Note that the hereditary
(tail) integral in (\ref{eq:4.32}) involves a logarithmic kernel. A
priori, one should include in the logarithm a constant time scale {\it
P} in order to adimensionalize the integration variable $\lambda$, say
$\ln (\lambda / 2P)$. However, $\ln P$ would actually be in factor of
an instantaneous term [depending only on the current instant $t$
through the sixth time-derivative $I^{(6)}_{ij}(t)]$, so
(\ref{eq:4.32}) is in fact independent of the choice of time scale. In
(\ref{eq:4.32}) we have chosen for simplicity $P=1$ sec.  The presence
of the tail integral (\ref{eq:4.32}) in the metric implies a
modification of the radiation reaction force at the relative 1.5PN
order \cite{BD88}. The other 4PN terms are not controlled at this
stage, but are instantaneous and thus do not yield any radiation
reaction effects (indeed the 4PN approximation is ``even" in the
post-Newtonian sense). It was further shown
\cite{BD92} that the 1.5PN tail integral in the radiation reaction is
such that there is exact energy balance with
a corresponding integral present in the far-zone
flux. Here we recover this fact and add it up to the results obtained
previously.

As the gauge transformation yielding (\ref{eq:4.32}) in \cite{BD88}
deals only with 4PN terms, it can be applied to the inner metric
$g^{\rm in}_{\mu \nu}$ given by (\ref{eq:3.1}) without modifying any
of the known terms at the 1PN non-radiative and reactive
approximations. It is clear from (\ref{eq:4.32}) and the reactive
potentials (\ref{eq:3.6}) that after gauge transformation, the inner
metric takes the same form as (\ref{eq:3.1}), except that the reactive
potentials are now more accurate, and given by

\begin{mathletters}
\label{eq:4.33}
\begin{eqnarray}
V^{\rm reac}({\bf x}, t) &=& -{G \over 5c^5} x_{ij} I^{(5)}_{ij} (t) +
{G \over c^7}
\left[{1\over 189}x_{ijk} I^{(7)}_{ijk}(t) - {1\over 70} {\bf x}^2 x_{ij}
I^{(7)}_{ij}(t) \right] \nonumber \\
&& - {4G^2M \over 5c^8}x_{ij} \int^{+\infty}_0
d\lambda \ln \left(\lambda\over 2 \right) I^{(7)}_{ij}(t-\lambda)
+ O\left(1\over c^9 \right) \, \label{eq:4.33a},\\
V^{\rm reac}_i ({\bf x}, t) &=& {G\over c^5}\left[{1\over 21} \hat x_{ijk}
I^{(6)}_{jk}(t)
- {4\over 45} \varepsilon_{ijk}x_{jm} J^{(5)}_{km}(t) \right] + O\left({1\over c^7}
\right)\, \label{eq:4.33b}.
\end{eqnarray}
\end{mathletters}
Still there remain in the metric some un-controlled (even) 4PN terms,
but these are made of {\it instantaneous} spatial integrals over the
source variables, exactly like the un-controlled 2PN and 3PN terms.
[The expressions (\ref{eq:4.33}) can be recovered also from
Sect.~III.D of paper I and a matching similar to the one performed in
this paper.]  With (\ref{eq:4.33}) in hands, one readily extends the
balance equations to 1.5PN order.  First one obtains (4.24), but where
the reactive potentials are given more accurately by (\ref{eq:4.33}),
and where there are some instantaneous 4PN terms $_8X$, $_8Y_i$ and
$_8Z_i$ in the right-hand-sides. Extending the (slight) assumption
made before concerning the similar 3PN terms, we can transform $_8X$,
$_8Y_i$ and $_8Z_i$ into time-derivatives and transfer them to the
left-hand-sides. This yields (4.25), except that the remainders in the
right-hand-sides are $O(c^{-9})$ instead of $O(c^{-8})$. Using
(\ref{eq:4.33}), we then obtain (working modulo total
time-derivatives) the laws (4.28) augmented by the tail contributions
arising at order $c^{-8}$ in the right-hand-sides. The remainders in
the left-hand-sides are of the order $d \widehat{O} (c^{-4})/dt =
O(c^{-9})$ (arguing as previously), and therefore are negligible as
compared to the tail contributions at $c^{-8}$. In the case of energy
the 1.5PN balance equation is obtained as
\begin{eqnarray}
{dE^{\rm 1PN}\over dt} &=& -{G\over 5c^5} I^{(3)}_{ij} I^{(3)}_{ij}
-{G\over c^7} \left[{1\over 189} I^{(4)}_{ijk} I^{(4)}_{ijk}
+{16 \over 45} J^{(3)}_{ij} J^{(3)}_{ij} \right] \nonumber \\
&& - {4G^2M \over 5c^8} I^{(3)}_{ij}(t) \int^{+\infty}_0 d\lambda \ln
\left(\lambda \over 2\right) I^{(5)}_{ij}(t-\lambda)+O\left({1\over c^9}
\right)\, \label{eq:4.34}.
\end{eqnarray}
Because there are no terms of order $c^{-3}$ in the internal energy
of the system (see Sect. IV.A), the energy $E^{1PN}$ appearing
in the left-hand-side is in fact valid at the 1.5PN order. Finally,
to the required order, one can re-write (\ref{eq:4.34}) equivalently
in a form where the flux of energy is manifestly positive-definite,
\begin{eqnarray}
{dE^{\rm 1PN}\over dt} &=& -{G\over 5c^5} \left({I^{(3)}_{ij}(t)+{2GM\over c^3}
\int^{+\infty}_0 d\lambda \ln \left(\lambda \over 2 \right) I^{(5)}_{ij}
(t-\lambda)} \right)^2 \nonumber \\ && - {G\over c^7} \left[{1\over
189} (I^{(4)}_{ijk})^2 +{16\over 45}(J^{(3)}_{ij})^2
\right]+O\left({1\over c^9}\right) \ .
\label{eq:4.35}
\end{eqnarray}
Under the latter form one recognizes in the right-hand-side the known
energy flux at 1.5PN order. Indeed the effective quadrupole moment
which appears in the parenthesis agrees with the tail-modified {\it
radiative} quadrupole moment parametrizing the field in the far zone
(see Eq. (3.10) in \cite{BD92}).  [The term associated with the
(gauge-dependent) constant 11/12 in the radiative quadrupole moment
\cite{BD92} yields a total time-derivative in the energy flux (as
would yield any time scale $P$ in the logarithm), and can be neglected
in (\ref{eq:4.35}).]  The 1.5PN balance equation for angular momentum
is proved similarly (it involves as required the same tail-modified
radiative quadrupole moment). The balance equation for linear momentum
does not include any tail contribution at 1.5PN order, and simply
remains in the form (\ref{eq:4.30}).

\acknowledgments

The author would like to thank Bala Iyer and Clifford Will for
stimulating discussions, and for their computation of the 1PN
radiation reaction potentials in the case of two-body systems
\cite{IW95}.  This prior computation clarified several issues, and
made easier the problem addressed in the present paper.  The author
would like also to thank Thibault Damour for discussions at an early
stage of this work, notably on the law of conservation of energy at
1PN order (Sect.~IV.A).  Finally, the author is very grateful to a
referee for his valuable remarks which have motivated an improved
version of the paper.


\begin{references}
\bibitem{B93}L. Blanchet, Phys. Rev. D{\bf 47}, 4392 (1993) [referred in
the text to as paper I].
\bibitem{IW93}B.R. Iyer and C.M. Will, Phys. Rev. Lett. {\bf 70}, 113 (1993).
\bibitem{IW95}B.R. Iyer and C.M. Will, Phys. Rev. D{\bf 52}, 6882 (1995).
\bibitem{HT75}R.A. Hulse and J.H. Taylor, Astrophys. J. (Letters)
{\bf 195}, L51 (1975).
\bibitem{TFMc79}J.H. Taylor, L.A. Fowler and P.M. Mc Culloch,
Nature {\bf 277}, 437 (1979).
\bibitem{D83b}T. Damour, Phys. Rev. Lett. {\bf 51}, 1019 (1983).
\bibitem{TWoDW}J.H. Taylor, A. Wolszczan, T. Damour and
J.M.  Weisberg, Nature {\bf 355}, 132 (1992).
\bibitem{T93}J.H. Taylor, Class. Quantum Grav. {\bf 10}, S167 (1993).
\bibitem{3mn}C. Cutler, T.A. Apostolatos, L. Bildsten, L.S. Finn,
E.E.~Flanagan, D.~Kennefick, D.M.~Markovic, A.~Ori, E.~Poisson,
G.J.~Sussman and K.S.~Thorne, Phys. Rev. Lett. {\bf 70}, 2984 (1993).
\bibitem{FCh93}L.S. Finn and D.F. Chernoff, Phys. Rev. D{\bf 47},
2198 (1993).
\bibitem{CF94}C. Cutler and E. Flanagan, Phys. Rev. D{\bf 49},
2658 (1994).
\bibitem{P93}E. Poisson, Phys. Rev. D{\bf 47}, 1497 (1993).
\bibitem{CFPS93}C. Cutler, L.S. Finn, E. Poisson and G.J. Sussmann,
Phys.  Rev.  D{\bf 47}, 1511 (1993).
\bibitem{TNaka94}H. Tagoshi and T. Nakamura, Phys. Rev. D{\bf
49}, 4016 (1994).
\bibitem{Sasa94}M. Sasaki, Prog. Theor. Phys. {\bf 92}, 17 (1994).
\bibitem{TSasa94}H. Tagoshi and M. Sasaki, Prog. Theor. Phys.
{\bf 92}, 745 (1994).
\bibitem{P95}E. Poisson, Phys. Rev. D{\bf 52}, 5719 (1995).
\bibitem{BDI95}L. Blanchet, T. Damour and B.R. Iyer,
Phys. Rev. D{\bf 51}, 5360 (1995).
\bibitem{BDIWW95}L. Blanchet, T. Damour, B.R. Iyer, C.M.~Will
and A.G. Wiseman, Phys. Rev. Lett. {\bf 74}, 3515 (1995).
\bibitem{WWi96}C.M. Will and A.G. Wiseman, Phys. Rev. D (in press).
\bibitem{B96pn}L. Blanchet, Phys. Rev. D{\bf 54}, 1417 (1996).
\bibitem{D83a}T. Damour, in {\it Gravitational Radiation}, N. Deruelle
and T. Piran (eds.), (North-Holland publishing Company, 1983), p. 59.
\bibitem{Bu69}W.L. Burke, unpublished Ph. D. Thesis, California
Institute of Technology (1969).
\bibitem{Th69}K.S. Thorne, Astrophys. J. {\bf 158}, 997 (1969).
\bibitem{Bu71}W.L. Burke, J. Math. Phys. {\bf 12}, 401 (1971).
\bibitem{C69}S. Chandrasekhar, Astrophys. J. {\bf 158}, 45 (1969).
\bibitem{CN69}S. Chandrasekhar and Y. Nutku, Astrophys. J.
{\bf 158}, 55 (1969).
\bibitem{CE70}S. Chandrasekhar and F.P. Esposito, Astrophys. J.
{\bf 160}, 153 (1970).
\bibitem{EhlRGH}J. Ehlers, A. Rosenblum, J.N. Goldberg and P.
Havas, Astrophys. J. {\bf 208}, L77 (1976).
\bibitem{WalW80}M. Walker and C.M. Will, Astrophys. J. {\bf 242},
L129 (1980).
\bibitem{BD84}L. Blanchet and T. Damour, Phys. Lett. {\bf 104}A, 82
(1984).
\bibitem{AD75}J.L. Anderson and T.C. DeCanio, Gen. Relat. Grav.
{\bf 6}, 197 (1975).
\bibitem{PaL81}A. Papapetrou and B. Linet, Gen. Relat. Grav. {\bf 13},
335 (1981).
\bibitem{Ehl80}J. Ehlers, Ann. N.Y. Acad. Sci. {\bf 336}, 279 (1980).
\bibitem{Ker80}G.D. Kerlick, Gen. Rel. Grav. {\bf 12}, 467 (1980).
\bibitem{Ker80'}G.D. Kerlick, Gen. Rel. Grav. {\bf 12}, 521 (1980).
\bibitem{BRu81}R. Breuer and E. Rudolph, Gen. Rel. Grav. {\bf 13},
777 (1981).
\bibitem{BRu82}R. Breuer and E. Rudolph, Gen. Rel. Grav. {\bf
14}, 181 (1982).
\bibitem{S85}G. Sch\"afer, Ann. Phys. (N.Y.) {\bf 161}, 81 (1985).
\bibitem{DD81a}T. Damour and N. Deruelle, Phys. Lett. {\bf 87A}, 81
(1981).
\bibitem{DD81b}T. Damour and N. Deruelle, C. R. Acad. Sc. Paris
{\bf 293}, 537 (1981).
\bibitem{D82}T. Damour, C. R. Acad. Sc. Paris, {\bf 294}, 1355 (1982).
\bibitem{BD86}L. Blanchet and T. Damour, Philos. Trans. R. Soc.
London A{\bf 320}, 379 (1986).
\bibitem{B87}L. Blanchet, Proc. R. Soc. Lond. A {\bf 409}, 383 (1987).
\bibitem{BD92}L. Blanchet and T. Damour, Phys. Rev. D{\bf 46}, 4304
(1992).
\bibitem{Bo59}W.B. Bonnor, Philos. Trans. R. Soc. London A{\bf 251},
233 (1959).
\bibitem{BoR66}W.B. Bonnor and M.A. Rotenberg, Proc. R. Soc. London
A{\bf 289}, 247 (1966).
\bibitem{HR69}A.J. Hunter and M.A. Rotenberg, J. Phys. A{\bf 2},
34 (1969).
\bibitem{Th80}K.S. Thorne, Rev. Mod. Phys. {\bf 52}, 299 (1980).
\bibitem{BD89}L. Blanchet and T. Damour, Ann. Inst. H. Poincar\'e (Phys.
Th\'eorique) {\bf 50}, 377 (1989).
\bibitem{DI91a}T. Damour and B.R. Iyer, Ann. Inst. H. Poincar\'e
(Phys.  Th\'eorique) {\bf 54}, 115 (1991).
\bibitem{B95}L. Blanchet, Phys. Rev. D{\bf 51}, 2559 (1995).
\bibitem{BD88}L. Blanchet and T. Damour, Phys. Rev. D{\bf 37},
1410 (1988).
\bibitem{BoR61}W.B. Bonnor and M.A. Rotenberg, Proc. R. Soc. London
A{\bf 265}, 109 (1961).
\bibitem{Pa71}A. Papapetrou, Ann. Inst. H. Poincar\'e XIV, 79 (1971).
\bibitem{Bek73}J.D. Bekenstein, Astrophys. J. {\bf 183}, 657 (1973).
\bibitem{Press77}W.H. Press, Phys. Rev. D{\bf 15}, 965 (1977).
\bibitem{Peres62}A. Peres, Phys. Rev. {\bf 128}, 2471 (1962).
\bibitem{N}Our conventions and notation are the following~: signature
$-+++$~; greek indices =0,1,2,3; latin indices =1,2,3; $g={\rm det}$
$(g_{\mu\nu})$; $\eta_{\alpha\beta}= \eta^{\alpha\beta}$ =~flat metric
=~diag (-1,1,1,1);
$r=|{\bf x}|=(x_1^2 +x_2^2 +x_3^2)^{1/2}$; $n^i =n_i =x^i/r$;
$\partial_i =\partial/\partial x^i$; $x^L =x_L= x_{i_1} x_{i_2}\ldots
x_{i_l}$ and $\partial_L =\partial_{i_1} \partial_{i_2}\ldots
\partial_{i_l}$, where $L=i_1 i_2\ldots i_l$ is a multi-index with $l$
indices; $x_{L-1} =x_{i_1}...x_{i_{l-1}}$, $x_{aL-1} =x_a x_{i_1} \ldots
x_{i_{l-1}}$, etc\dots; $\hat x_L$ and $\hat\partial_L$ are the
(symmetric) and trace-free parts of $x_L$ and $\partial_L$, for instance
$\hat x_{ij} =x_ix_j-{1\over 3} \delta_{ij} r^2$;
$\hat x_L$ is denoted also $x_{<L>}$;
$T_{(ij)} = {1\over 2} (T_{ij} + T_{ji})$; the superscript $(n)$ denotes
$n$ time-derivatives.
\bibitem{S83}G. Sch\"afer, Lett. Nuovo Cim. {\bf 36}, 105 (1983).
\bibitem{Fock}V.A. Fock, {\it Theory of Space, Time and
Gravitation}, Pergamon, London (1959).
\bibitem{R}In the appendix B of Ref.\cite{B95} the expression
of the 1PN-conserved energy $E^{\rm 1PN}$ as given by the equations (B3)
and (B7) is in error. The correct expression is (4.18) in the present
paper. The agreement between the equations (B3) and (B7)
was proved in Ref.~\cite{B95} only when both equations take their
correct form.
\bibitem{EW75}R. Epstein and R.V. Wagoner, Astrophys. J. {\bf 197},
717 (1975).
\bibitem{Th83}K.S. Thorne, in {\it Gravitational Radiation}, N. Deruelle
and T. Piran (eds.), (North-Holland publishing Company, 1983), p. 3.
\end{references}
\end{document}